\documentclass[prd,preprint,showpacs,groupedaddress,floatfix,showkeys]{revtex4-1}
\usepackage{amsmath}
\usepackage{amsfonts}
\usepackage{amssymb}
\usepackage{geometry}
\usepackage[english]{babel}
\usepackage{graphicx}
\usepackage{subfigure}
\usepackage{indentfirst}
\usepackage{mathrsfs}
\usepackage{overpic}
\usepackage{hyperref}
\begin{document}

  \setlength{\parindent}{2em}
  \title{Shadows of rotating Hayward-de Sitter black holes with astrometric observables}
  \author{Peng-Zhang He} 
 \author{Qi-Qi Fan}
 \author{Hao-Ran Zhang}
  \author{Jian-Bo Deng} \email[Jian-Bo Deng: ]{dengjb@lzu.edu.cn}
  \affiliation{Institute of Theoretical Physics $\&$ Research Center of Gravitation, Lanzhou University, Lanzhou 730000, China}
  \date{\today}

  \begin{abstract}
Motivated by recent work on rotating black hole shadow [Phys. Rev. D101, 084029 (2020)], we investigate the shadow behaviors of rotating Hayward-de Sitter black hole for static observers at a finite distance in terms of astronomical observables. This paper uses the newly introduced distortion parameter in [arXiv:2006.00685] to describe the shadow's shape quantitatively. We show that the spin parameter would distort shadows and the magnetic monopole charge would increase the degree of deformation. At the same time, the distortion could be relieved because of the cosmological constant and the distortion would increase with the distance from the black hole. Besides, the spin parameter, magnetic monopole charge and cosmological constant increase will cause the shadow to shrink.

  \end{abstract}

  %\pacs{04.20.-q, 04.50.-h}


  \maketitle

  \section{Introduction}
  According to General Relativity (GR), the most interesting celestial body predicted may be black holes. There is a strong gravitational field in the region near a black hole that can bend light rays. Due to the highly bending light rays in the strong gravity field, shadows cast by black holes usually appear in the observer's sky~\cite{1}. The light rays received come from the black hole's unstable photon orbits, or the photon region~\cite{2,3}. The first image of a black hole taken by Event Horizon Telescope (EHT)~\cite{4} confirmed the existence of black holes, attracting more researchers to study the observable effects of the black holes, e.g., the shadows of black holes, gravitational deflection of light or massive particles and the like.
  \par
  For the simplest black hole, spherical black hole, the shadow's boundary is a perfect circle. In the sixties of the last century, Synge considered a static observer to calculate the angular radius of the Schwarzschild black hole's shadow in his seminal paper~\cite{synge}. For rotating black holes, the shadow's shape is no longer circular but somewhat flattened on one side because of the ``dragging" of null geodesics by black holes. Bardeen first gave the shadow's shape of the Kerr black hole for a distant observer; one can find the results in Chandrasekhar's book~\cite{Chandrasekhar} and in~\cite{bardeen}. Since those pioneer works, shadows of objects have been extensively studied; one can refer to the Refs.~\cite{perlick,changzhe1,cunha,Badia,Konoplya,Abdujabbarov,Atamurotov,Bisnovatyi-Kogan,Younsi,Abdujabbarov2,Tsupko,Papnoi,Cunha1,cunha2,Atamurotov1,Amir,Sharif,Babar,Kumar,Singh}.
  \par
  Very recently, the authors of Refs.~\cite{changzhe2,changzhe3} proposed a new method for calculating the size and shape of shadow in terms of astrometric observables for finite-distant observers and introduced a new distortion parameter to describe the shadow's deviation from circularity. The shadows of the Kerr-de Sitter black hole for static observers were revisited in this way without introducing tetrads~\cite{changzhe2}. Furthermore, the appearance of the shadow of a static spherical black hole and the Kerr black hole was discussed in a unified framework~\cite{changzhe3}. 
  \par
    This paper aims to apply this method to study the shadows of rotating Hayward-de Sitter black holes and examine the parameters' effects on the shadow's size and distortion.
  \par
  We organize this article as follows. In Sec.~\ref{sec2}, we review the method of calculating black hole shadows using astronomical observables briefly. In sec.~\ref{sec3}, we apply this approach to rotating Hayward-de Sitter black holes to analyze the influences of parameters on the shadow's shape and size. We conclude our results in Sec.~\ref{sec4}. In this paper we set $ G=c=1 $.

  \section{SHADOWS OF rotating BLACK HOLES}\label{sec2}
 In order to make this article self-sufficient, we briefly introduce some basics in this section. One can read Refs.~\cite{changzhe2,changzhe3} for details.
 \par
  In astrometry, the angle $ \epsilon $ between two incident light rays can be expressed by the following formula~\cite{Lebedev}:
  \begin{equation}\label{eq:angle1}
  \cos \epsilon \equiv \frac{\gamma^{*} w \cdot \gamma^{*} k}{\left|\gamma^{*} w \| \gamma^{*} k\right|}=\frac{w \cdot k}{(u \cdot w)(u \cdot k)}+1.
  \end{equation}
   Here, $ k $ and $ w $ are tangent vectors of the two light rays, respectively. $ \gamma^{*} $ is the projection operator, $ \gamma_{\nu}^{\mu}=\delta_{\nu}^{\mu}+u^{\mu} u_{\nu} $, for a given observer, whose 4-velocity is denoted by vector $ u $. 
  \par
  Generally speaking, the metric of a rotating black hole can be written as  
  \begin{equation}
  \mathrm{d} s^{2}=g_{00} \mathrm{d} t^{2}+g_{11} \mathrm{d} r^{2}+g_{22} \mathrm{d} \theta^{2}+g_{33} \mathrm{d} \phi^{2}+2 g_{03} \mathrm{d} t \mathrm{d} \phi.
  \end{equation}
  The 4-velocity of a static observer is $ u=\frac{1}{\sqrt{g_{00}}} \partial_{t} $. For the asymptotically de Sitter spacetime, there is a cosmological horizon. The observer is fixed at the so-called domain of outer communication that is the region between the event horizon and the cosmological horizon. When the observer located at $ \theta =0 $, it will find that the shadow is a disk and the angular radius is 
  \begin{equation}\label{eq7}
  \cot \psi =\operatorname{sgn}\left(\frac{\pi}{2}-\psi\right) \sqrt{\frac{g_{11}}{g_{22}\left(\frac{l^{2}}{l^{1}}\right)^{2}+\left(g_{33}-\frac{g_{03}^{2}}{g_{00}}\right)\left(\frac{l^{3}}{l^{2}}\right)^{2}}}.
  \end{equation}
  Here, we have choose a light ray $ l=\left(l^{0},l^{1},l^{2},l^{3}\right) $ comes from the photon region. ``$ \text{sgn} $" represents the sign function. For a observer located at $ \theta>0 $, the shadow's silhouette is not a perfect circle as a consequence of the frame dragging effect. As an example, assume the observer located at $ \theta =\pi/2 $. Let $ k=\left(k^{0},k^{1},0,k^{3}\right) $  represent a light ray from a prograde  orbit which moves in the same direction as the black hole's rotation, and $ w=\left(w^{0},w^{1},0,w^{3}\right) $ represent a light ray from a retrograde  orbit that moving against the black hole's rotation. One can get the angle of the two light rays, in such a way that 
  \begin{equation}\label{8}
  \cot\gamma=\operatorname{sgn}(k, w) \sqrt{\frac{\left(\frac{g_{11}}{\mathcal{K}-\mathcal{W}}+\left(g_{33}-\frac{g_{03}^{2}}{g_{00}}\right) \frac{1}{\frac{1}{\mathcal{W}}-\frac{1}{\mathcal{K}}}\right)^{2}}{g_{11}\left(g_{33}-\frac{g_{03}^{2}}{g_{00}}\right)}},
  \end{equation}
  where $\mathcal{K}\equiv k^{3}/k^{1},\mathcal{W} \equiv w^{3}/w^{1}  $, and $ \operatorname{sgn}(k, w)=\operatorname{sgn}\left(\cos\gamma\right)=\operatorname{sgn}\left(g_{11}+\left(g_{33}-g_{03}^{2}/g_{00}\right) \mathcal{K} \mathcal{W}\right) $.
  
  \par
  Similarly, the angle $ \alpha  $ between a light ray $ l=\left(l^{0},l^{1},l^{2},l^{3}\right) $ from the photon region and $ k $ is
  \begin{equation}\label{10}
  \cot \alpha  = {\rm{sgn}}(k,l)\sqrt {\frac{{{{\left( {{g_{11}}\frac{1}{{\mathcal{K} - {\mathcal{L}_3}}} + \left( {{g_{33}} - \frac{{g_{03}^2}}{{{g_{00}}}}} \right)\frac{1}{{\frac{{\rm{1}}}{{{\mathcal{L}_3}}} - \frac{1}{\mathcal{K}}}}} \right)}^2}}}{{{g_{22}}\left( {{g_{11}}{{\left( {\frac{{{\mathcal{L}_2}}}{{\mathcal{K} - {\mathcal{L}_3}}}} \right)}^2} + \left( {{g_{33}} - \frac{{g_{03}^2}}{{{g_{00}}}}} \right){{\left( {\frac{{{\mathcal{L}_2}}}{{1 - \frac{{{\mathcal{L}_{\rm{3}}}}}{\mathcal{K}}}}} \right)}^2}} \right) + {g_{11}}\left( {{g_{33}} - \frac{{g_{03}^2}}{{{g_{00}}}}} \right)}}} ;
  \end{equation}
  and the angle $ \beta  $ between the light ray $ l $ and $ w $ is
  \begin{equation}\label{11}
  \cot \beta  = {\mathop{\rm sgn}} (w,l)\sqrt {\frac{{{{\left( {{g_{11}}\frac{1}{{\mathcal{W} - {\mathcal{L}_3}}} + \left( {{g_{33}} - \frac{{g_{03}^2}}{{{g_{00}}}}} \right)\frac{1}{{\frac{{\rm{1}}}{{{\mathcal{L}_3}}} - \frac{1}{\mathcal{W}}}}} \right)}^2}}}{{{g_{22}}\left( {{g_{11}}{{\left( {\frac{{{\mathcal{L}_2}}}{{\mathcal{W} - {\mathcal{L}_3}}}} \right)}^2} + \left( {{g_{33}} - \frac{{g_{03}^2}}{{g_{00}^2}}} \right){{\left( {\frac{{{\mathcal{L}_2}}}{{1 - \frac{{{\mathcal{L}_3}}}{{{\mathcal{W}}}}}}} \right)}^2}} \right) + {g_{11}}\left( {{g_{33}} - \frac{{g_{03}^2}}{{{g_{00}}}}} \right)}}}  .
  \end{equation}
  In above equations, $ \mathcal{L}_{2} \equiv l^{2}/l^{1}, \mathcal{L}_{3} \equiv l^{3}/l^{1} $, $ \operatorname{sgn}(k, l)=\operatorname{sgn}\left(\cos\alpha \right)=\operatorname{sgn}\left(g_{11}+\left(g_{33}-g_{03}^{2}/g_{00}\right) \mathcal{K} \mathcal{L}_{3}\right) $,  and $  \operatorname{sgn}(w, l)=\operatorname{sgn}\left(\cos\beta \right)=\operatorname{sgn}\left(g_{11}+\left(g_{33}-g_{03}^{2}/g_{00}\right) \mathcal{W}\mathcal{L}_{3} \right) $.
  \par
  $ \gamma  $, $ \alpha  $ and $ \beta $ can provide us the shadow of black hole in the celestial sphere. For the sake of convince in researching the shadow, one can use the following stereographic projection for the celestial coordinates to describe the shape of shadow in a $ 2D $-plane~\cite{changzhe2}.
  \begin{equation}\label{14}
  \begin{aligned}
  Y_{\mathrm{sh}}&=\frac{2 \sin \Phi \sin \Psi}{1+\cos \Phi \sin \Psi} \\
  &=\frac{2 \cos \beta \sin \gamma-2 \cot \gamma \sqrt{\sin ^{2} \gamma \sin ^{2} \beta+(\cos (\beta+\gamma)-\cos \alpha)(\cos (\beta-\gamma)-\cos \alpha)}}{1+\cos \beta \cos \gamma+\sqrt{\sin ^{2} \gamma \sin ^{2} \beta+(\cos (\beta+\gamma)-\cos \alpha)(\cos (\beta-\gamma)-\cos \alpha)}},
  \end{aligned}
  \end{equation}
  \begin{equation}\label{15}
  \begin{aligned}
  Z_{\mathrm{sh}}&=\frac{2 \cos \Psi}{1+\cos \Phi \sin \Psi}\\
  &=\frac{2 \csc \gamma \sqrt{(\cos \alpha-\cos (\beta+\gamma))(\cos (\beta-\gamma)-\cos \alpha)}}{1+\cos \beta \cos \gamma+\sqrt{\sin ^{2} \gamma \sin ^{2} \beta+(\cos (\beta+\gamma)-\cos \alpha)(\cos (\beta-\gamma)-\cos \alpha)}}.
  \end{aligned}	
  \end{equation}
Here, $ \Phi $ and $ \Psi $ are azimuth angle and polar angle in celestial coordinate system.
  \par
  In order to quantitatively describe shadow's shape, a distortion parameter $ \Xi $ in terms of $ \alpha  $, $ \beta $ and $ \gamma $ is introduced,
  \begin{equation}
  \cos \Xi \equiv \frac{1+\cos \gamma-\cos \alpha-\cos \beta}{2 \sqrt{(1-\cos \alpha)(1-\cos \beta)}},
  \end{equation}
  where $ \Xi $ ranges from $ 0 $ to $ \pi $. The $ \cos\Xi =0 $ for that shadow's shape is circular in the celestial sphere. For the non-vanished $ \cos\Xi $, it can quantify the deviation from circularity. The authors of Ref.~\cite{changzhe3} first proposed this kind of quantity for the shadow. Now, we can use $ \gamma $ and $ \Xi $ to represent the sizes and shapes of shadows without confusion.

  \section{APPLICATION IN ROTATING HAYWARD-DE SITTER BLACK HOLES}\label{sec3}
  In this section, we will apply this method described in the previous section to obtain the shadows of rotating Hayward-de Sitter black holes without introducing tetrads.
  \par
  The metric of rotating Hayward-ds Sitter black holes in the Boyer-Lindquist coordinates $ \left(t,r,\theta,\phi\right) $ is~\cite{Ali}
   \begin{equation}\label{eq16}
   	d s^{2}=-\frac{\Delta_{r}}{\Sigma}\left(d t-\frac{a \sin ^{2} \theta}{\rho} d \phi\right)^{2}+\frac{\Sigma}{\Delta_{r}} d r^{2}+\frac{\Sigma}{\Delta_{\theta}} d \theta^{2}+\frac{\Delta_{\theta} \sin ^{2} \theta}{\Sigma}\left(a d t-\frac{r^{2}+a^{2}}{\rho} d \phi\right)^{2},
   \end{equation}
   where 
   \begin{equation}
   	\Sigma=r^2+a^{2}\cos^{2}\theta,\quad\rho=1+\frac{\Lambda}{3}a^{2},
   \end{equation}
   \begin{equation}
   	\Delta_{r}=\left(r^{2}+a^{2}\right)\left(1-\frac{\Lambda}{3}r^{2}\right)-2 \tilde{m}\left(r\right)r,\quad\Delta_{\theta}=1+\frac{\Lambda}{3}a^{2}\cos^{2}\theta,
   \end{equation}
   \begin{equation}
   	\tilde{m}\left(r\right)=M\left(\frac{r^{3}}{r^{3}+g^{3}}\right).
   \end{equation}
  Here, $ M $ represents the mass of black hole, $ a $ is the black hole spin parameter, $\Lambda  $ is cosmological constant, and the parameter $ g $ is the magnetic monopole charge arising from the nonlinear electrodynamics.

\subsection{Null geodesic equations and photon regions}
The motion equations of photons in the spacetime, determined by the metric~\eqref{eq16}, can be given by the Lagrangian,
\begin{equation}
	\mathcal{L}=\frac{1}{2}g_{\mu\nu}\dot{x}^{\mu}\dot{x}^{\nu},
\end{equation}  
where an overdot denotes the partial derivative with respect to an affine parameter. For the metric ~\eqref{eq16}, one can obtain the momenta ($ p_{\mu}=g_{\mu\lambda}\dot{x}^{\lambda} $) are  
\begin{gather}\label{eq21}
	p_{t}=\left(\frac{a^{2}\Delta_{\theta}\sin^{2}\theta}{\Sigma}-\frac{\Delta_{r}}{\Sigma}\right)\dot{t}+\left(\frac{a \Delta_{r} \sin ^{2} \theta}{\rho \Sigma}-\frac{a\left(a^{2}+r^{2}\right) \Delta_{\theta} \sin ^{2} \theta}{\rho \Sigma}\right)\dot{\phi},\\
	p_{\phi}=\left(\frac{a \Delta_{r} \sin ^{2} \theta}{\rho \Sigma}-\frac{a\left(a^{2}+r^{2}\right) \Delta_{\theta} \sin ^{2} \theta}{\rho \Sigma}\right)\dot{t}+\left(\frac{\left(a^{2}+r^{2}\right)^{2} \Delta_{\theta} \sin ^{2} \theta}{\rho^{2} \Sigma}-\frac{a^{2} \Delta_{r} \sin ^{4} \theta}{\rho^{2} \Sigma}\right)\dot{\phi}\label{eq22},\\
	p_{r}=\frac{\Sigma}{\Delta_{r}}\dot{r},\\ p_{\theta}=\frac{\Sigma}{\Delta_{\theta}}\dot{\theta},\label{eq23}
\end{gather}
where $ p_{t}=-E$, $p_{\phi}=L_{\phi} $ denote energy and angular momentum, respectively. Combining the momenta and Hamilton-Jacobi equation, we can get null geodesics equations.
\par
The Hamilton-Jacobi equation takes the following general form: 
\begin{equation}\label{eq24}
	-\frac{\partial S}{\partial \lambda}=\frac{1}{2} g^{\mu\nu}\frac{\partial S}{\partial x^{\mu}}\frac{\partial S}{\partial x^{\nu}},
\end{equation}
where $ \lambda $ is an affine parameter and $ S $ is the Jacobi action which can be decomposed as a sum,
\begin{equation}\label{eq25}
	S=-\frac{1}{2}m^{2}\lambda-Et+L_{\phi}\phi+S_{\theta}\left(\theta\right)+S_{r}\left(r\right),
\end{equation}
if $ S $ is separable. $ m $ is the mass of particle, which is zero for photons. From~\eqref{eq24} and~\eqref{eq25}, one can get 
\begin{equation}
	\Delta_{\theta}\left(\frac{\partial S_{\theta}}{\partial \theta}\right)^{2}+\frac{\left(L_{\phi}\rho\csc\theta-aE\sin\theta\right)^{2}}{\Delta_{\theta}}=\mathcal{Q},
\end{equation}
and 
\begin{equation}
	\Delta_{r}\left(\frac{\partial S_{r}}{\partial r}\right)^{2}-\frac{\left(\left(a^{2}+r^{2}\right)E-a\rho L_{\phi}\right)^{2}}{\Delta_{r}}=-\mathcal{Q},
\end{equation}
where $ \mathcal{Q} $ is a constant of separation called Carter constant, and $ \partial S/\partial x^{\mu}=p_{\mu} $. With the Hamilton-Jacobi equation, it is not difficult to get the null geodesic equations as
\begin{gather}
	\left(\Sigma \dot{r}\right)^{2}=R,\\
	 (\Sigma \dot{\theta})^{2}=\Theta,\label{28}\\
	\Sigma \dot{t}=E\left(\frac{\left(a^{2}+r^{2}\right)\left(a^{2}+r^{2}-a \lambda \rho\right)}{\Delta_{r}}+\frac{a\left(\lambda \rho-a \sin ^{2} \theta\right)}{\Delta_{\theta}}\right),\\
	\Sigma \dot{\phi}=\rho E\left(\frac{a\left(a^{2}+r^{2}\right)-a^{2} \lambda \rho}{\Delta_{r}}+\frac{\left(\lambda \rho-a \sin ^{2} \theta\right)}{\Delta_{\theta} \sin ^{2} \theta}\right),\label{30}
\end{gather}
where
\begin{gather}\label{31}
	R=E^{2}\left(\left(a^{2}+r^{2}-a \lambda \rho\right)^{2}-\eta \Delta_{r}\right),\\
	\Theta=E^{2}\left(\Delta_{\theta} \eta-(\lambda \rho \csc \theta-a \sin \theta)^{2}\right),\label{32}
\end{gather}
and 
\begin{equation}
	\lambda\equiv\frac{L_{\phi}}{E},\quad \eta\equiv\frac{\mathcal{Q}}{E^{2}}.
\end{equation}
For spherical orbits,
\begin{equation}
	R\left(r_{c}\right)=0
\end{equation}
and 
\begin{equation}
	\left.\frac{dR\left(r\right)}{dr}\right|_{r=r_{c}}=0
\end{equation}
must be satisfied, which lead to 
\begin{gather}
	\lambda=\left.\frac{-4 r \Delta_{r}+\left(a^{2}+r^{2}\right) \Delta_{r}^{\prime}}{a \rho \Delta_{r}^{\prime}}\right|_{r=r_{c}},\\
	\eta=\left.\frac{16 r^{2} \Delta_{r}}{\Delta_{r}^{\prime}\,^2}\right|_{r=r_{c}},
\end{gather}
where $ \Delta_{r}^{\prime} $ denotes the derivative of $ \Delta_{r} $ with respect to $ r $, and $ r_{c} $ is the location of photon sphere. Furthermore, we can rewrite $ R^{\prime\prime}\left(r_{c}\right) $ as
\begin{equation}\label{38}
	R^{\prime\prime}\left(r_{c}\right)=\left.8 E^{2}\left(r^{2}+\frac{2r\Delta_{r}\left(\Delta_{r}^{\prime}-r\Delta_{r}^{\prime\prime}\right)}{\Delta_{r}^{\prime}\,^{2}}\right)\right|_{r=r_{c}}.
\end{equation}
A spherical null geodesic at $ r=r_{c} $ is unstable with respect to radial perturbations if $ R^{\prime\prime}\left(r_{c}\right)>0 $, and stable if $ R^{\prime\prime}\left(r_{c}\right)<0. $ Unstable photon orbits determine the contour of shadow. The range of $ r_{c} $ (photon region) can be determined by $ \Theta\ge 0 $ from~\eqref{28} and~\eqref{32}, which is 
\begin{equation}\label{39}
	\left(\left(4 r \Delta_{r}-\Sigma \Delta_{r}^{\prime}\right)^{2}-16 a^{2} r^{2} \Delta_{r} \Delta_{\theta} \sin ^{2} \theta\right)_{r=r_{c}} \leq 0.
\end{equation} 
From~\eqref{38} and~\eqref{39}, we can get $r_{c-}\le r_{c}\le r_{c+}  $, where $ r_{c-} $ and $ r_{c+}$ are the minimum and maximum radial position of the photon region. If limiting the light rays from the photon region, one can regard $ p^{\mu}=\dot{x}^{\mu} $ as functions of $ x^{\mu} $, $E  $, and $ r_{c} $. 

\subsection{Sizes of shadow}
For $ \theta =0 $, one can rewrite~\eqref{39} as 
\begin{equation}
	\left(4 r \Delta_{r}-\left(r^{2}+a^{2}\right) \Delta_{r}^{\prime}\right)_{r=r_{c}}=0.
\end{equation}
This means that the photon region becomes photon sphere, and $ r_{c}=r_{c-}=r_{c+} $. Substituting the metric~\eqref{eq16} and geodesic equations into~\eqref{eq7}, one can calculate the angular radius of the shadow in the following form,
\begin{equation}
	\cot \psi=\sqrt{\frac{\left(a^{2}+r^{2}-a \lambda  \rho\right)^{2}-\eta \Delta_{r}}{\Delta_{r} \eta+a \lambda \rho\left(2 a^{2}+2 r^{2}-a \lambda \rho\right)}},
\end{equation}
where $ \lambda $ and $ \eta $ are function of $ r_{c} $. Here, we only consider shadow in the view of observers located outside of the photon region. 
\begin{center}
	\begin{figure}[ht]
		\centering
		\subfigure[]{
			\begin{minipage}[t]{0.311\textwidth}
				\includegraphics[width=1\textwidth]{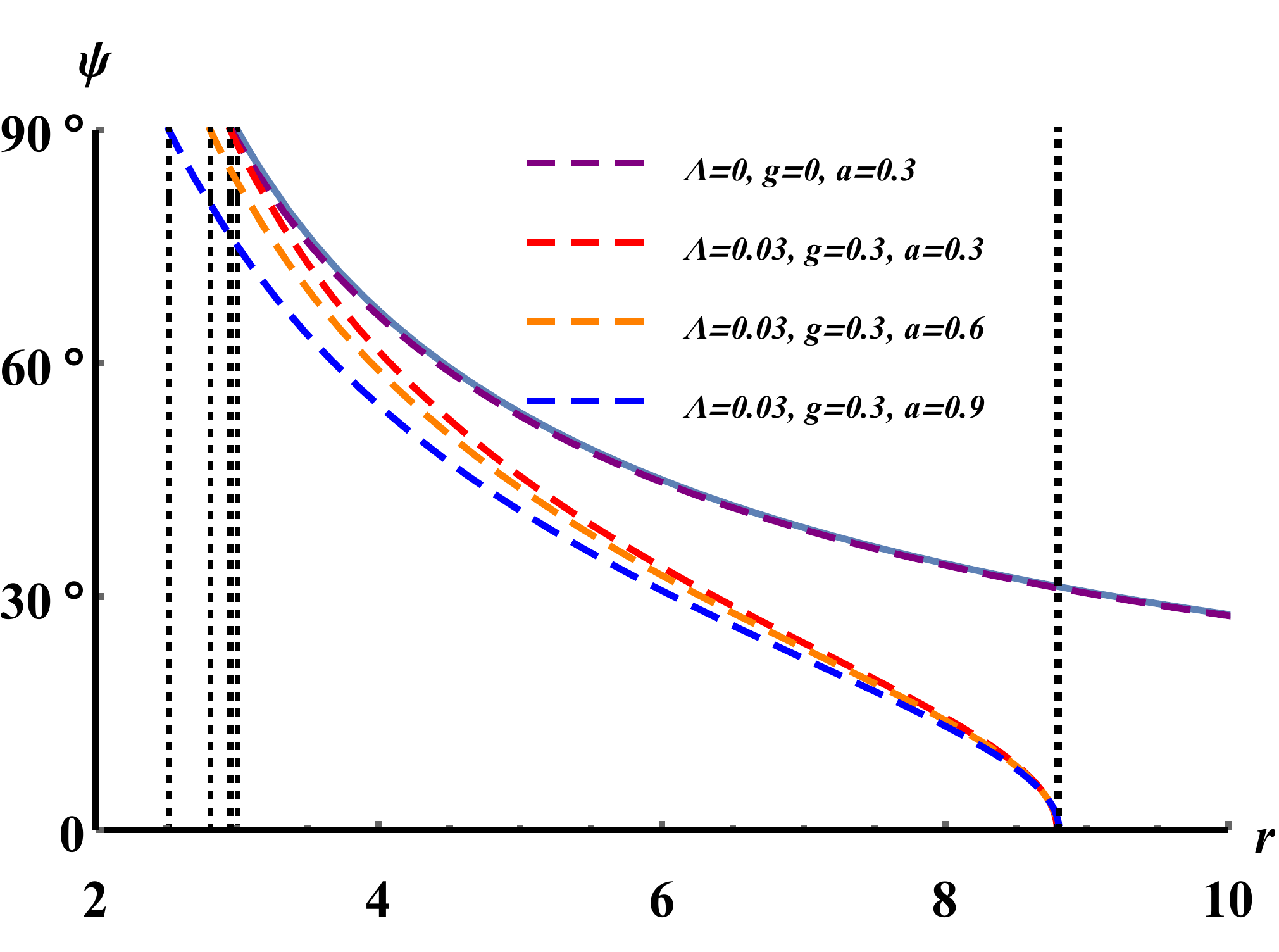}
			\end{minipage}\label{5a}
		}	
		\subfigure[]{
			\begin{minipage}[t]{0.311\textwidth}
				\includegraphics[width=1\textwidth]{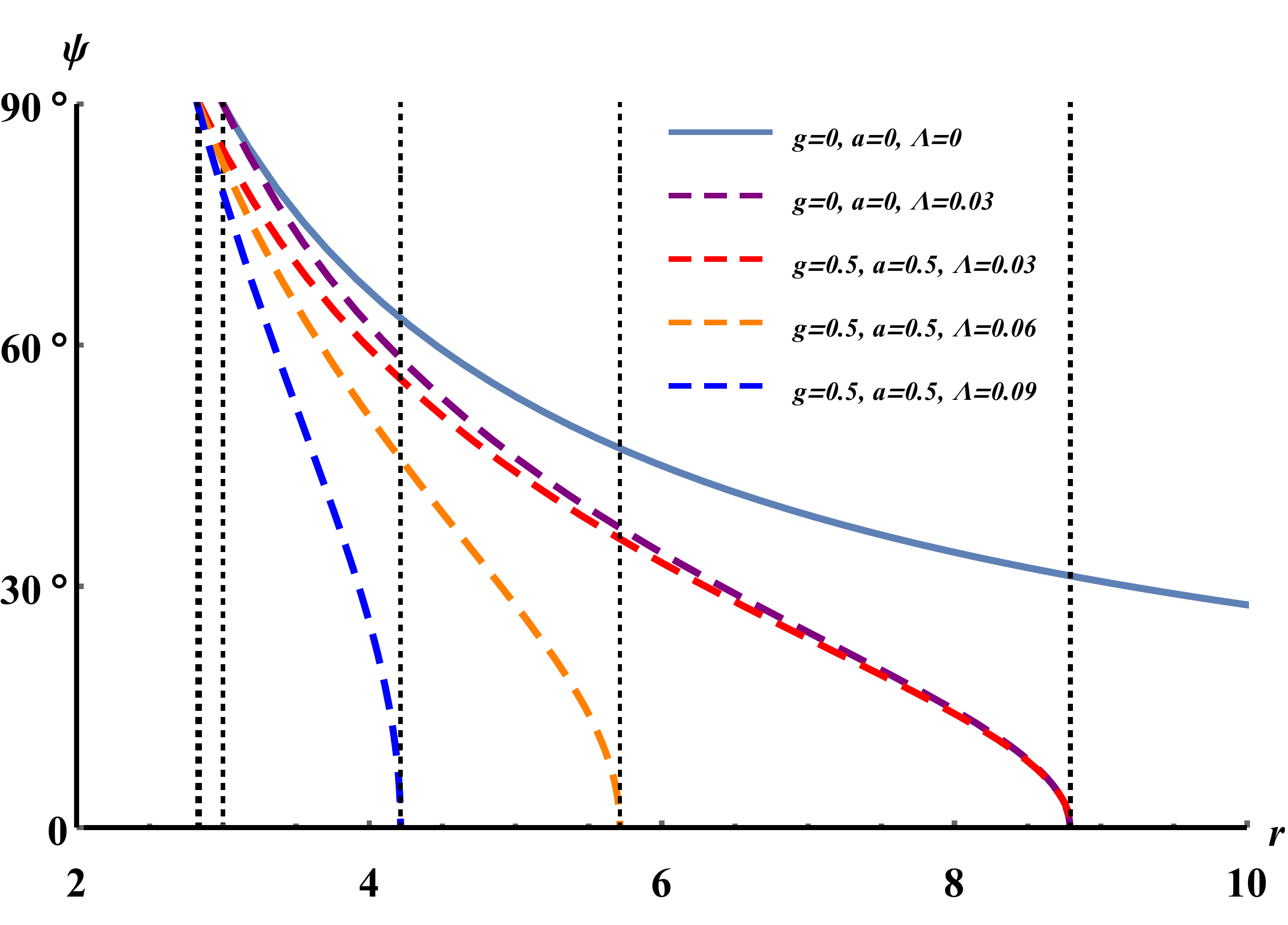}
			\end{minipage}\label{5b}
		}
		\subfigure[]{
			\begin{minipage}[t]{0.311\textwidth}
				\includegraphics[width=1\textwidth]{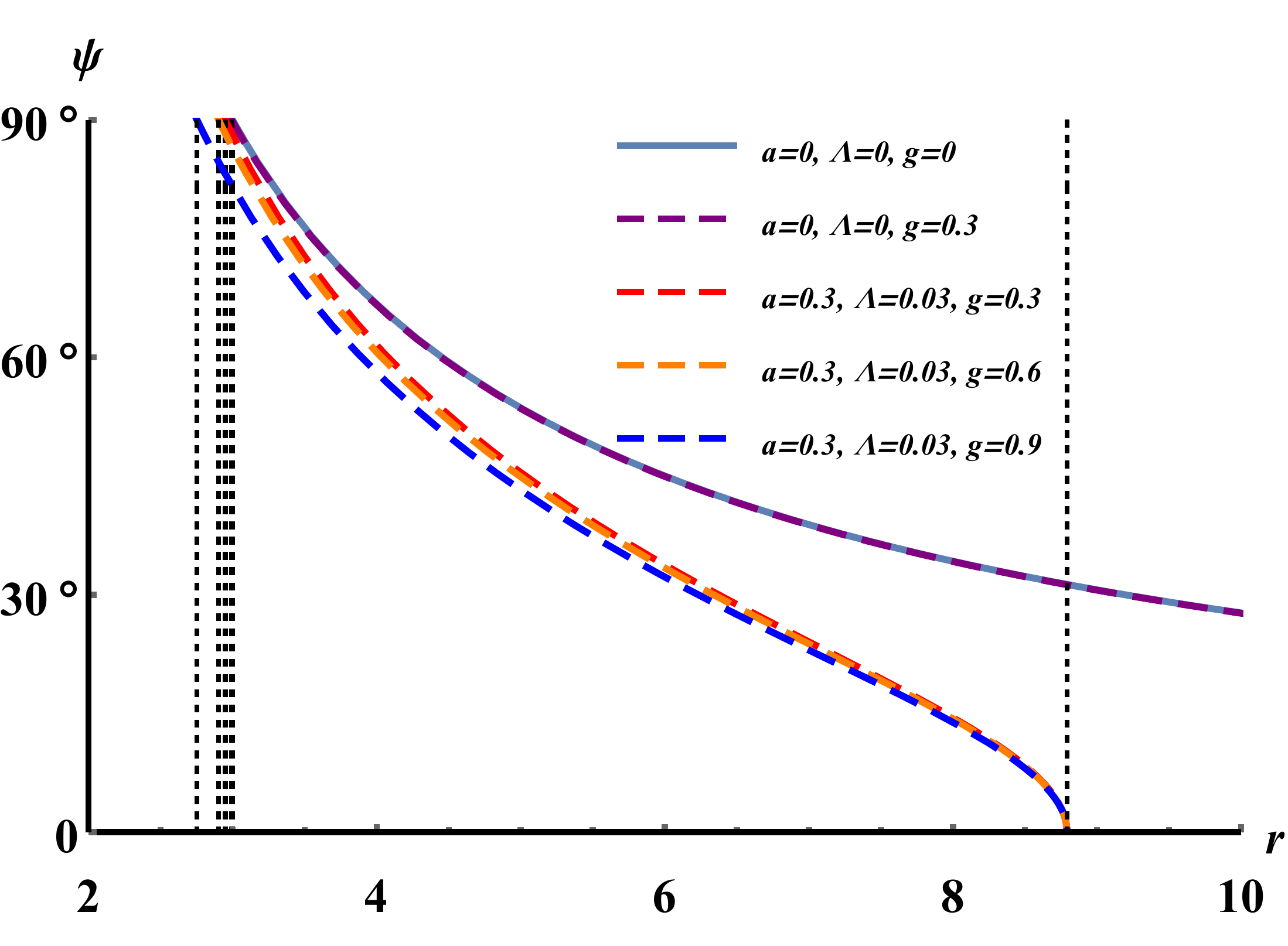}
			\end{minipage}\label{5c}
		}
		\caption{The angular radius $ \psi $ of shadow as a function of the distance from the rotating Hayward-de Sitter black holes for selected parameters, and the observers are located at inclination angle $ \theta=0 $. The vertical dotted lines are the outer boundaries and the cosmological horizons. Here we set $ M=1 $.}
		\label{fig:5}
	\end{figure}
\end{center}

In Fig.~\ref{fig:5}, we plot the shadow's angular radius as a function of the distance from the black hole. The figures reflect that the photon sphere radius of the Schwarzschild black hole is the largest, and it's shadow has the largest size among the shadows observed at the same position. Besides, no matter which of $ a $, $ g $, and $ l $ increases, the size of shadow will become smaller.
\par
\begin{center}
	\begin{figure}[htbp]
		\centering
		\subfigure[]{
			\begin{minipage}[t]{0.311\textwidth}
				\includegraphics[width=1\textwidth]{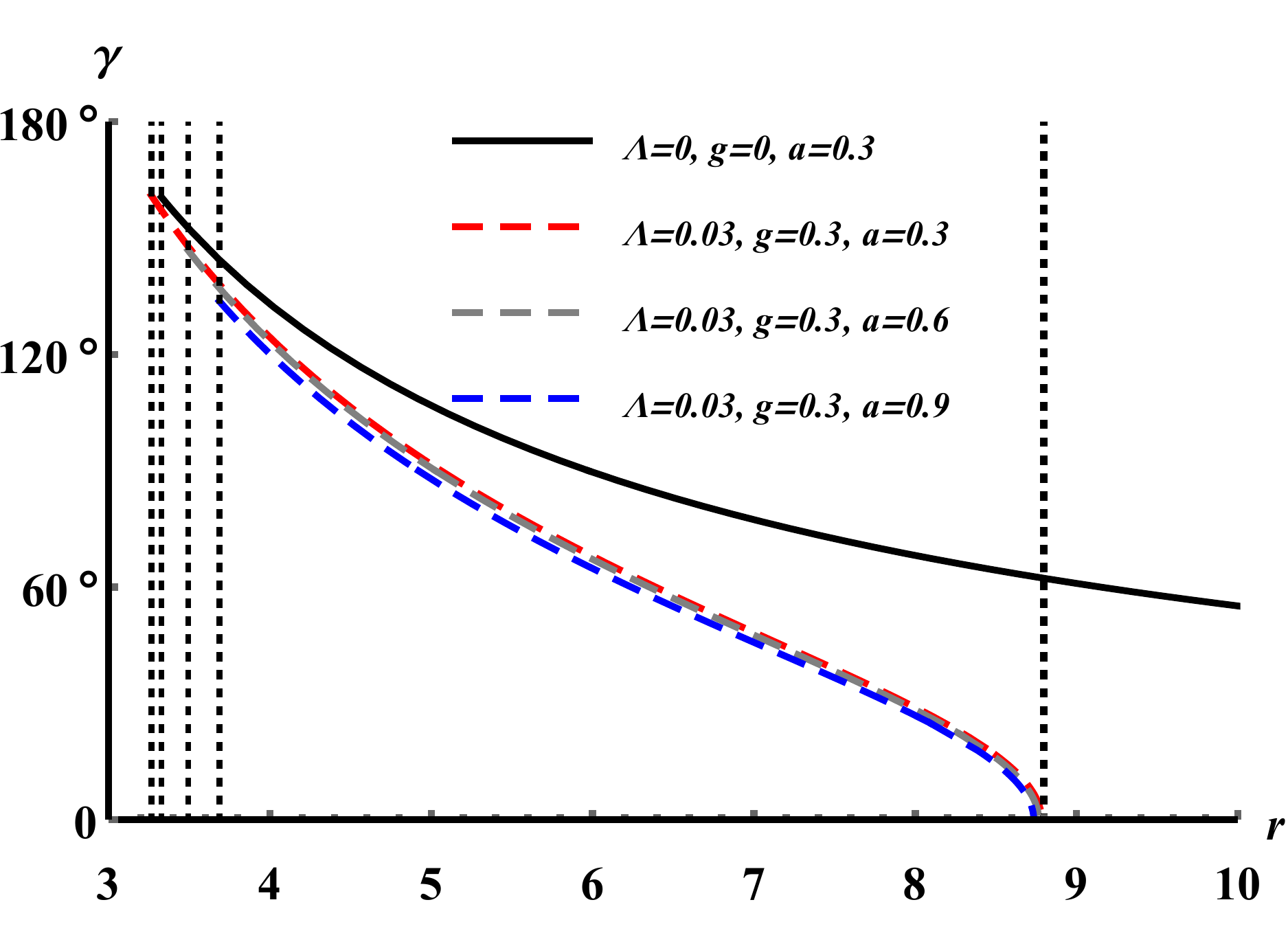}
			\end{minipage}\label{6a}
		}	
		\subfigure[]{
			\begin{minipage}[t]{0.311\textwidth}
				\includegraphics[width=1\textwidth]{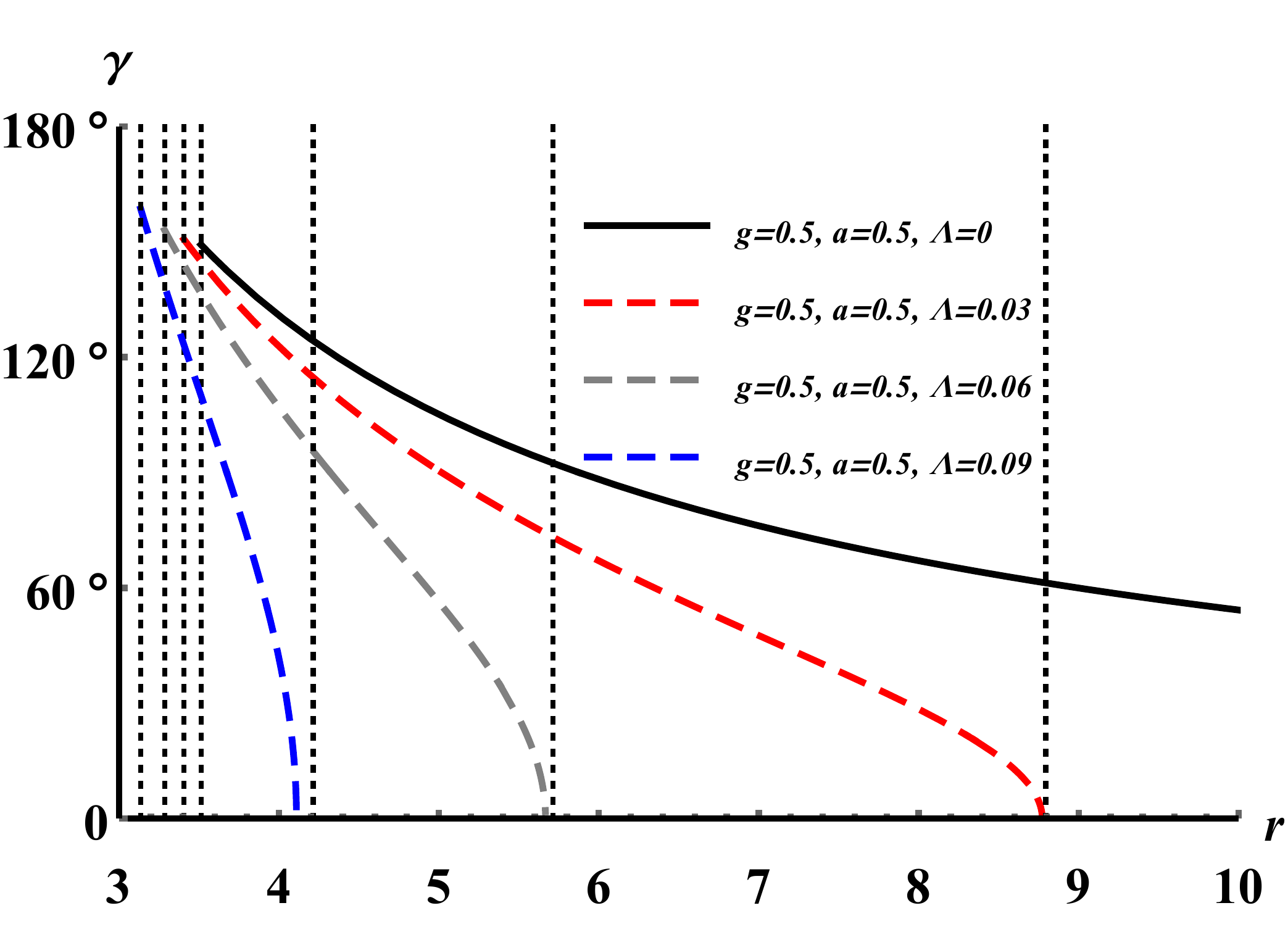}
			\end{minipage}\label{6b}
		}
		\subfigure[]{
			\begin{minipage}[t]{0.311\textwidth}
				\includegraphics[width=1\textwidth]{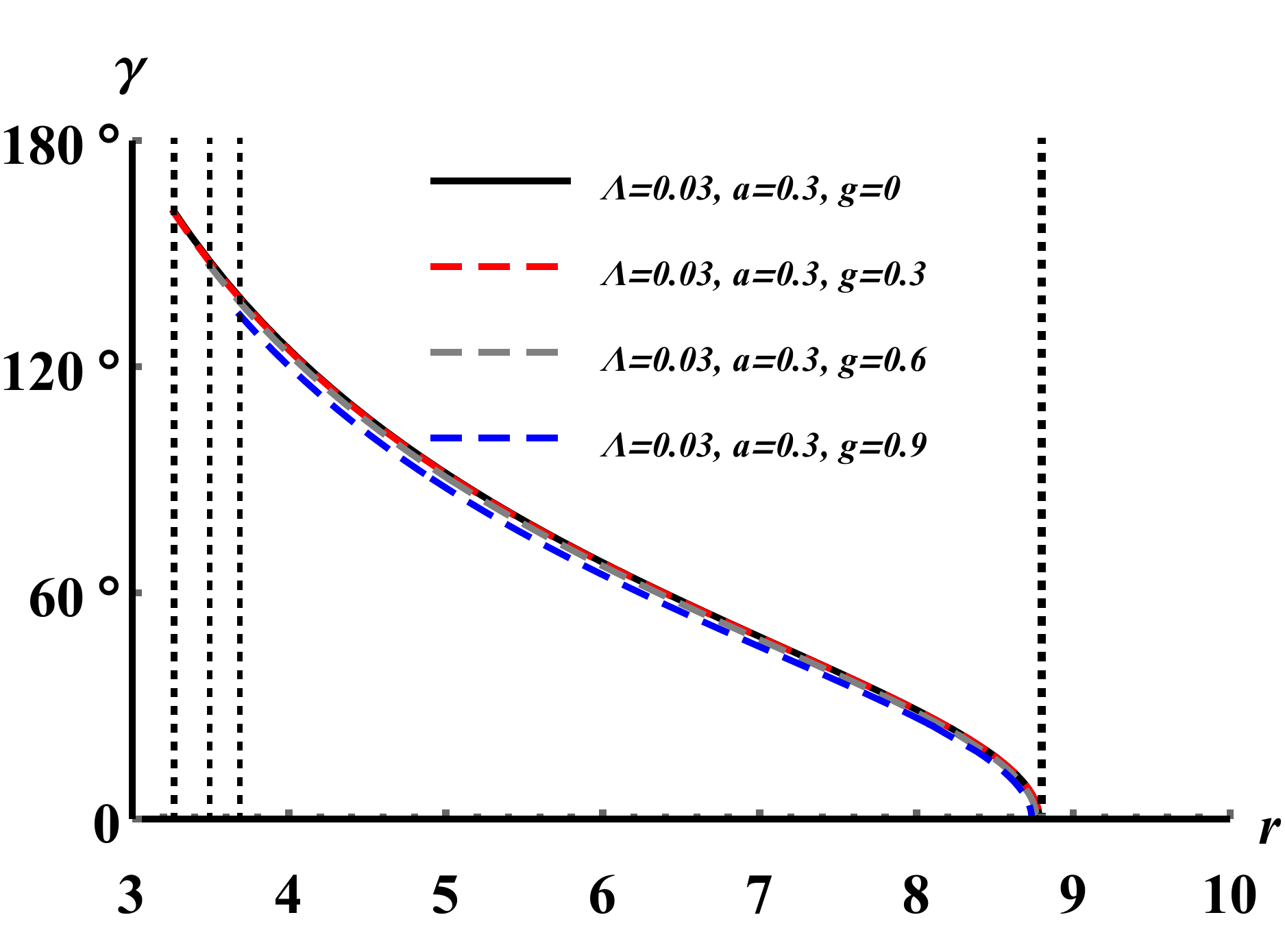}
			\end{minipage}\label{6c}
		}
		\caption{The angular diameter  $ \gamma  $ of shadow as a function of the distance from the rotating Hayward-de Sitter black holes for selected parameters, and the observers are located at inclination angle $ \theta=0 $. The vertical dotted lines are the outer boundaries and the cosmological horizons. Here we set $M=1.$}
		\label{fig:6}
	\end{figure}
\end{center}

The situation of the observer locates at the equatorial plane ($ \theta=\pi/2 $) will be more complicated. In this case,~\eqref{39} can be rewritten as 
\begin{equation}
	\left(\left(4 r \Delta_{r}-r^{2} \Delta_{r}^{\prime}\right)^{2}-16 a^{2} r^{2} \Delta_{r}\right)_{r=r_{c}} \leq 0.
\end{equation}
Then one can obtain $ r_{c-}\le r_{c}\le r_{c+} $. From~\eqref{8}, we get the angular diameter $ \gamma $,
\begin{equation}
	\cot \gamma=\operatorname{sgn}\left(1+\frac{\Delta_{r}^{2}}{\rho^{2}\left(\Delta_{r}-a^{2}\right)} \mathcal{K} \mathcal{W}\right) \left| \frac{\rho \sqrt{\Delta_{r}-a^{2}}}{\Delta_{r}} \frac{1}{\mathcal{K}-\mathcal{W}} +\frac{\Delta_{r}}{\rho \sqrt{\Delta_{r}-a^{2}} } \frac{1}{\frac{1}{\mathcal{W}}-\frac{1}{\mathcal{K}}}\right|,	
\end{equation} 
where
\begin{equation}\label{44}
	\mathcal{K}=\left.\frac{p^{\phi}}{p^{r}}\right|_{r_{c}=r_{c-}},
\end{equation}
\begin{equation}\label{45}
	\mathcal{W}=\left.\frac{p^{\phi}}{p^{r}}\right|_{r_{c}=r_{c+}}.
\end{equation}
From~\eqref{28} and~\eqref{30}, we get
\begin{equation}\label{46}
	\frac{p^{\phi}}{p^{r}}=\frac{\dot{\phi}}{\dot{r}}=\frac{\rho\left(a^{3}+a\left(r^{2}-\Delta_{r}\right)-a^{2} \lambda \rho+\Delta_{r} \lambda \rho\right)}{\Delta_{r} \sqrt{\left(a^{2}+r^{2}-a \lambda \rho\right)^{2}-\Delta_{r} \eta}}.
\end{equation}
It is worth noting that $ \lambda $ and $ \eta $ can be regarded as functions of $ r_{c} $, so~\eqref{46} is a function of $ r $ and $ r_{c} $. Fig.~\ref{fig:6} shows that the angular parameter $ \gamma $ changes with the increase of the distance from the black hole. It is not difficult to find that the angular parameter $ \gamma  $ decreases with the increase of $ a $, $ \Lambda $ and $ g $, and the outer boundary of the photon region $ r_{c+} $ is larger than the radius of the photon sphere in Schwarzschild spacetime. Therefore, the size of the black hole shadow will decrease with the increase of $ a $, $ \Lambda $ and $ g $, and the shadow of the Schwarzschild black hole has the largest size. 
\subsection{Shadow's shape}
In this part, we will consider the shadow's shape in different situations. The observers located at inclination angle $ \theta=0 $ would see the shadows as a perfect circle, while the observers located at $ \theta=\frac{\pi}{2} $ would find that the shadows are distorted. according to~\eqref{10} and~\eqref{11} ,the angular distances $ \alpha $ and $ \beta  $ can be read as  
\begin{equation}
	\cot \alpha=\operatorname{sgn}\left(1+\frac{\Delta_{r}^{2}}{\left(\Delta_{r}-a^{2}\right) \rho^{2}} \mathcal{K} \mathcal{L}_{3}\right) \frac{\left|\frac{\Delta_{r}}{\rho\sqrt{\Delta_{r}-a^{2}}} \frac{1}{\frac{1}{\mathcal{L}_{3}}-\frac{1}{\mathcal{K}}}+\frac{\rho\sqrt{\Delta_{r}-a^{2}}}{\Delta r} \frac{1}{\mathcal{K}-\mathcal{L}_{3}}\right|}{\sqrt{1+\left(\frac{\mathcal{L}_{2}}{1-\frac{\mathcal{L}_{3}}{\mathcal{K}}}\right)^{2} \Delta r+\frac{\left(\Delta r-a^{2}\right) \rho^{2}}{\Delta r}\left(\frac{\mathcal{L}_{2}}{\mathcal{K}-\mathcal{L}_{3}}\right)^{2}}},
\end{equation}
and
\begin{equation}
	\cot \beta=\operatorname{sgn}\left(1+\frac{\Delta_{r}^{2}}{\left(\Delta_{r}-a^{2}\right) \rho^{2}} \mathcal{W} \mathcal{L}_{3}\right) \frac{\left|\frac{\Delta_{r}}{\rho\sqrt{\Delta_{r}-a^{2}}} \frac{1}{\frac{1}{\mathcal{L}_{3}}-\frac{1}{\mathcal{W}}}+\frac{\rho\sqrt{\Delta_{r}-a^{2}}}{\Delta_{r}} \frac{1}{\mathcal{W}-\mathcal{L}_{3}}\right|}{\sqrt{1+\left(\frac{\mathcal{L}_{2}}{1-\frac{\mathcal{L}_{3}}{\mathcal{W}}}\right)^{2} \Delta_{r}+\frac{\left(\Delta_{r}-a^{2}\right) \rho^{2}}{\Delta_{r}}\left(\frac{\mathcal{L}_{2}}{\mathcal{W}-\mathcal{L}_{3}}\right)^{2}}},
\end{equation}
where $ \mathcal{K} $ and $ \mathcal{W} $ are given by~\eqref{44},~\eqref{45} and 
\begin{equation}
	\mathcal{L}_{2}\equiv\left.\frac{p^{\theta}}{p^{r}}\right|_{r_{c}},
\end{equation}
\begin{equation}
	\mathcal{L}_{3}\equiv\left.\frac{p^{\phi}}{p^{r}}\right|_{r_{c}},
\end{equation}
with 
\begin{equation}
	\frac{p^{\theta}}{p^{r}}=\frac{\dot{\theta}}{\dot{r}}=\pm \sqrt{\frac{\eta-(\lambda \rho-a)^{2}}{\left(a^{2}+r^{2}-a \lambda \rho\right)^{2}-\Delta_{r} \eta}}.
\end{equation}
\begin{center}
	\begin{figure}[htbp]
		\centering
		\subfigure[]{
			\begin{minipage}[t]{\textwidth}
				\includegraphics[width=0.45\textwidth]{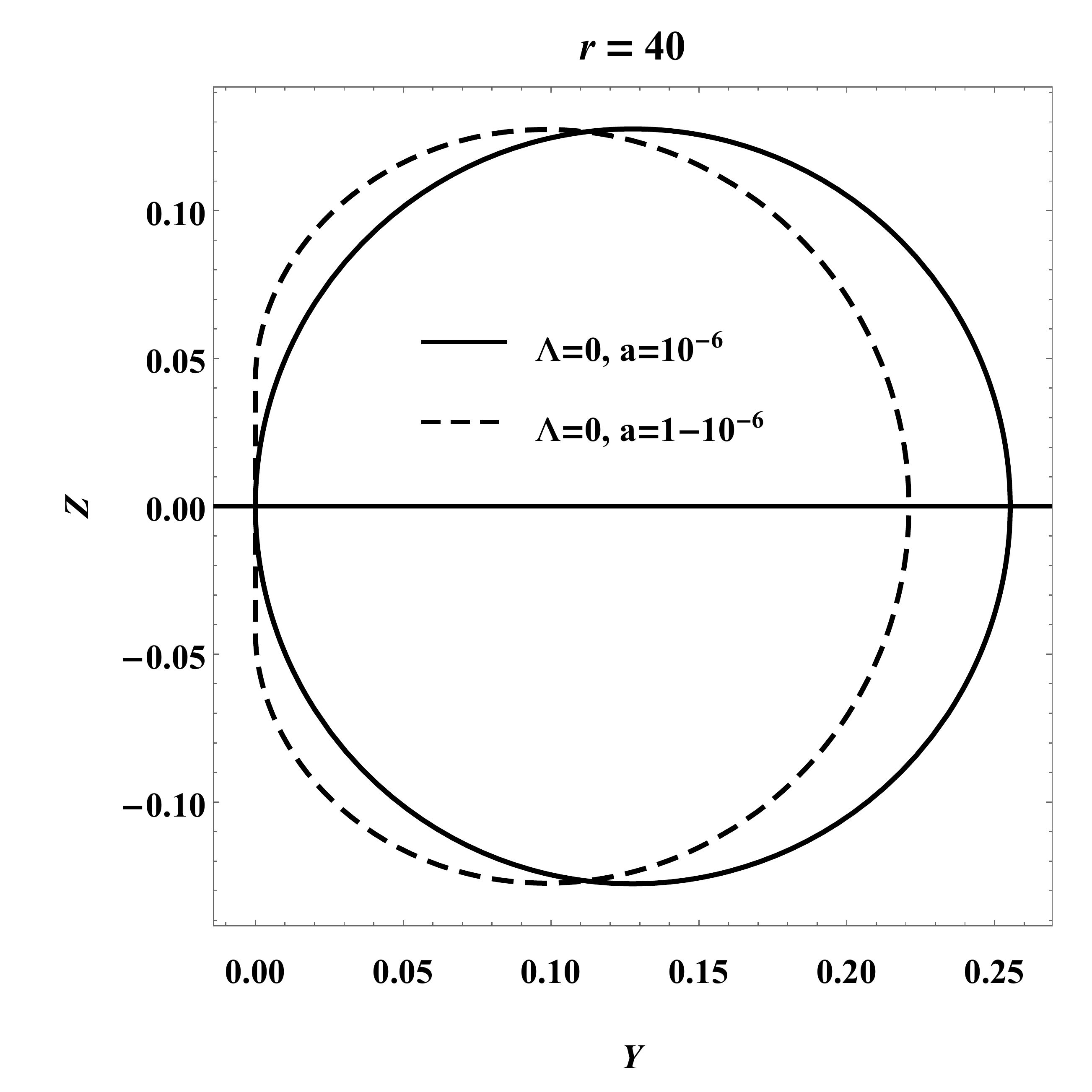}
				\includegraphics[width=0.45\textwidth]{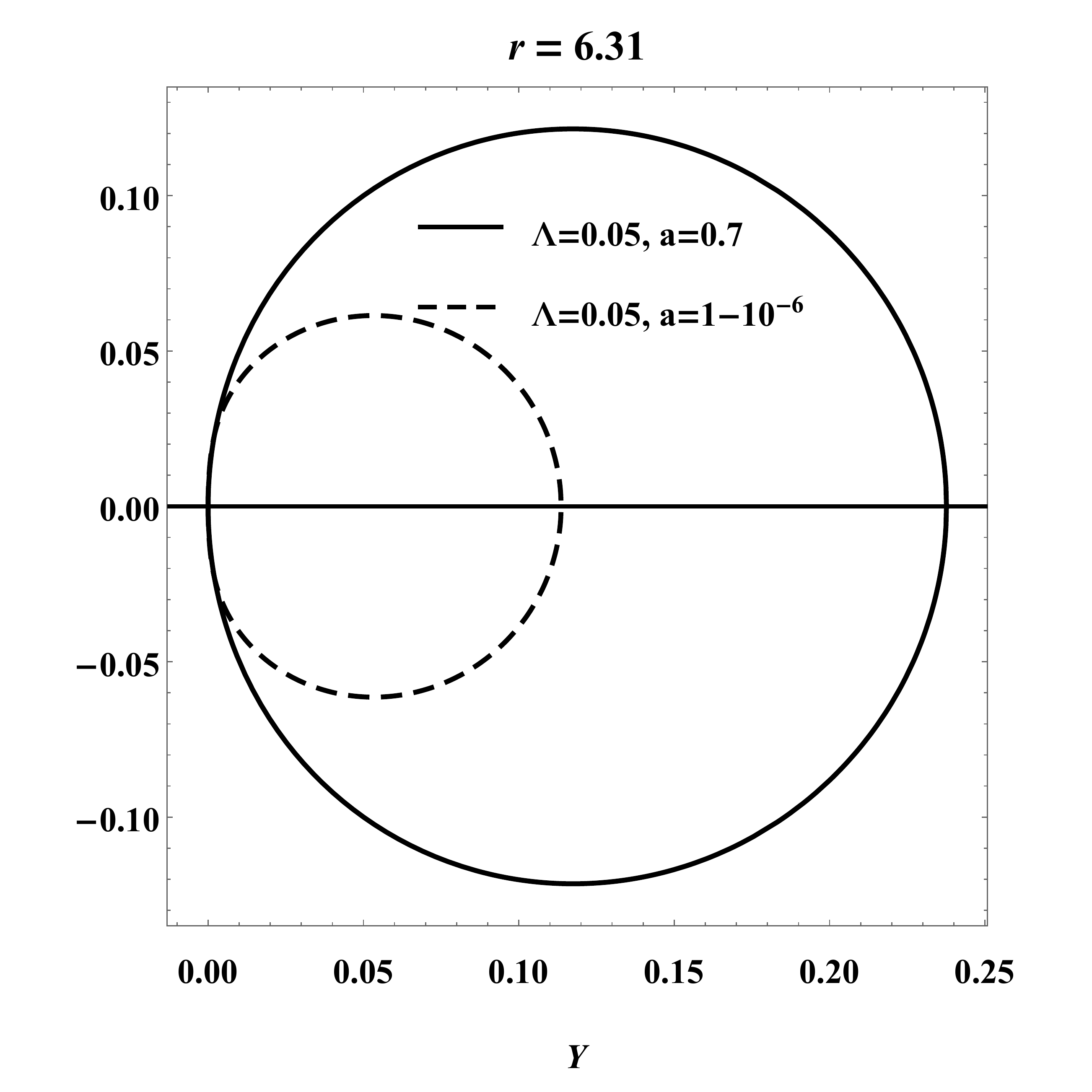}
			\end{minipage}
		}	
		\subfigure[]{
			\begin{minipage}[t]{\textwidth}
				\includegraphics[width=0.45\textwidth]{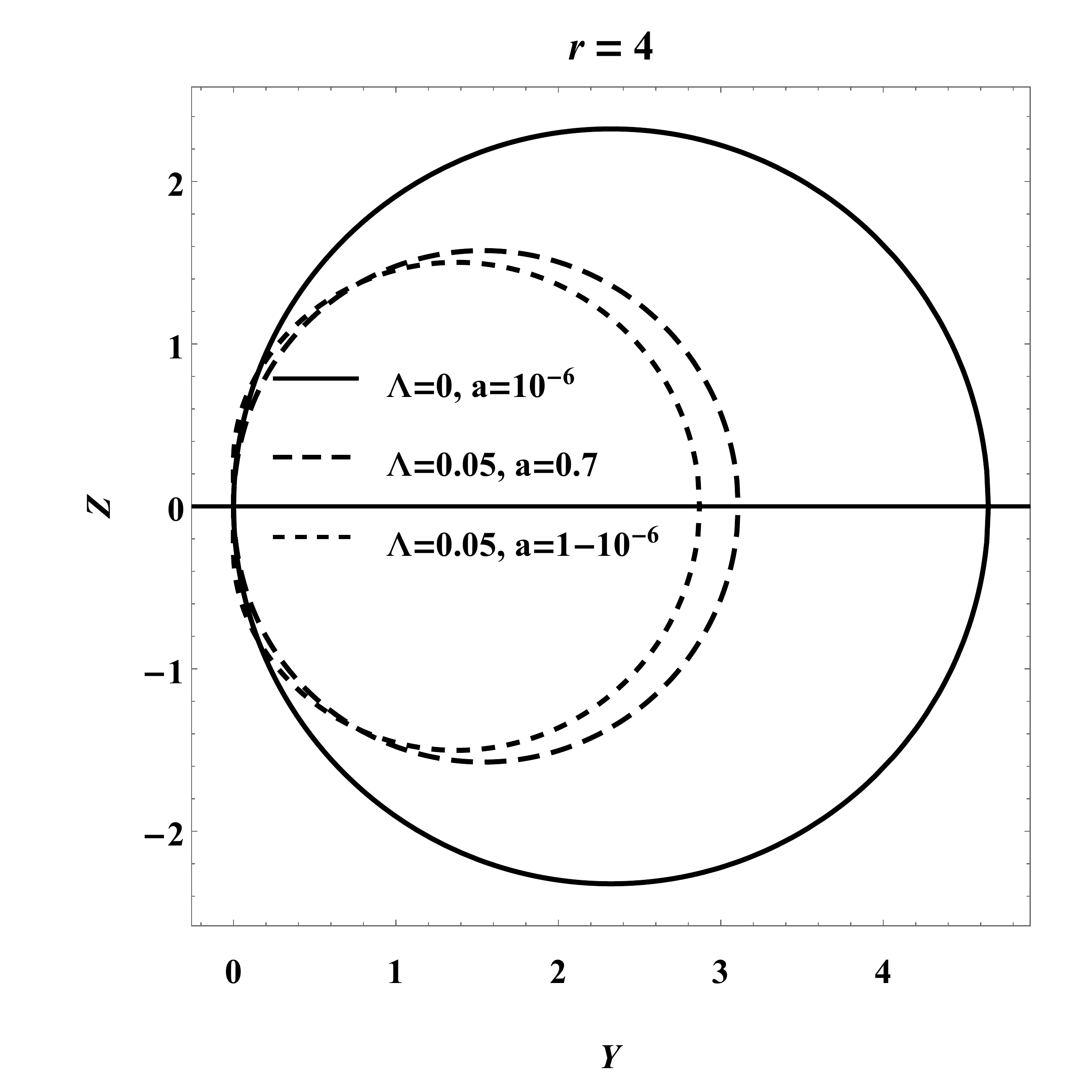}
				\includegraphics[width=0.45\textwidth]{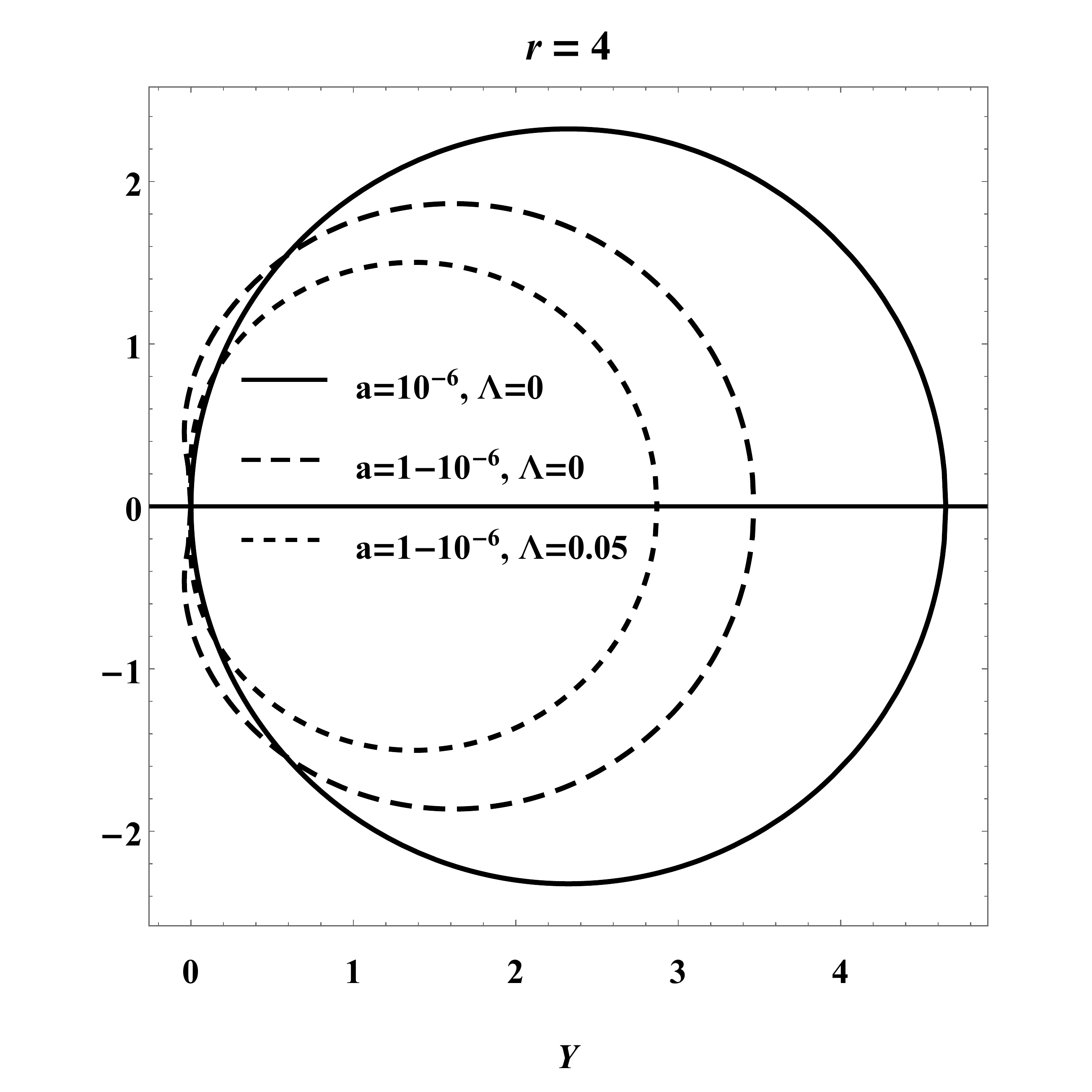}
			\end{minipage}
		}
		\caption{Shadows of rotating Hayward-de Sitter black holes with $ g=0 $ on projective plane $ \left(Y,Z\right) $ for selected parameters. $ r $ is the distance from the observer to the black holes. Here we set $ M=1 $. (a) Shadows of rotating Kerr(-de Sitter) black holes for selected spin parameters for distant observers. (b) Shadows of rotating Kerr-de Sitter black holes for observers located at $ r=4 $. }
		\label{fig:7}
	\end{figure}
\end{center}

	\begin{figure}[htbp]
		\centering
		\subfigure[]{
			\begin{minipage}[t]{0.311\textwidth}
				\includegraphics[width=1\textwidth]{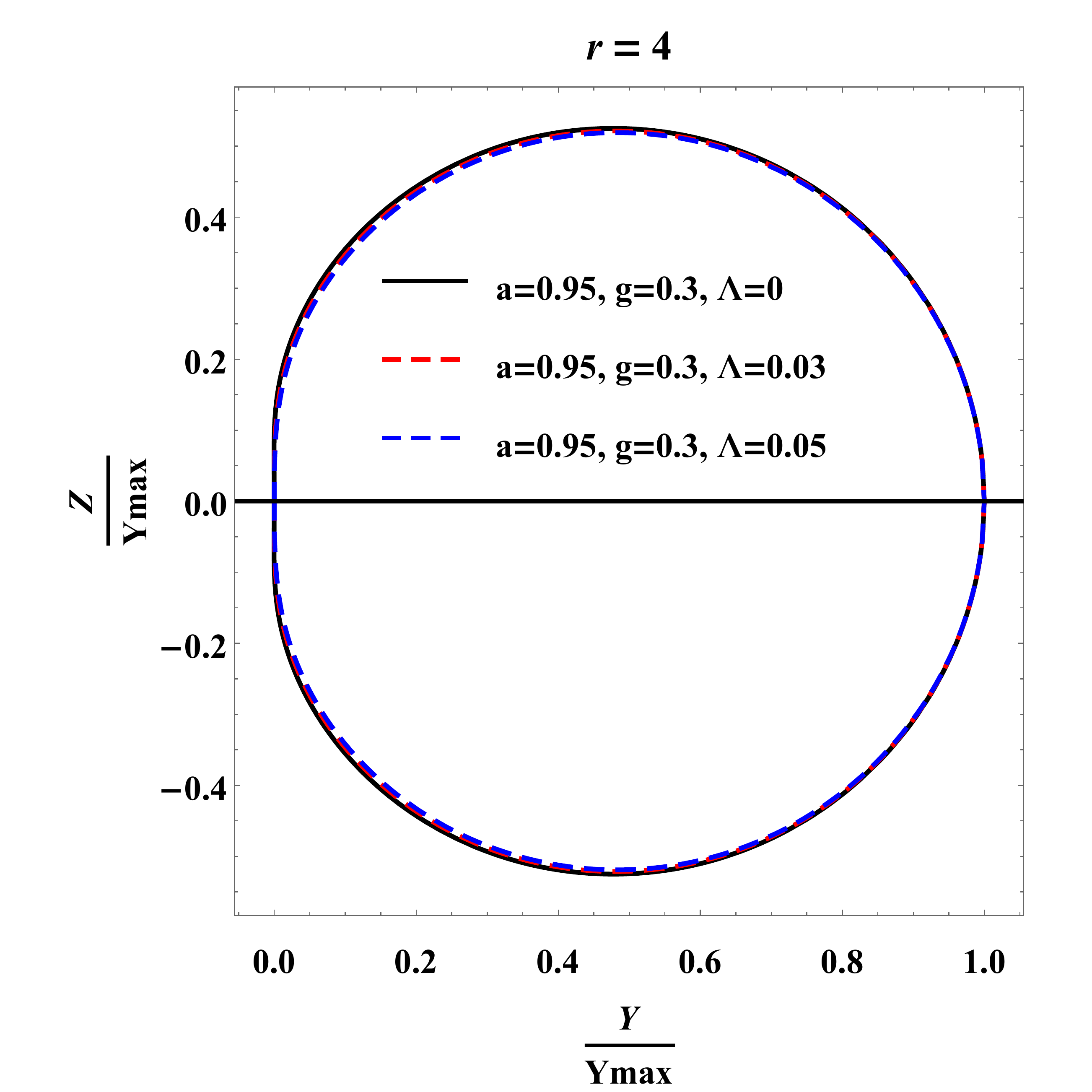}
				\includegraphics[width=1\textwidth]{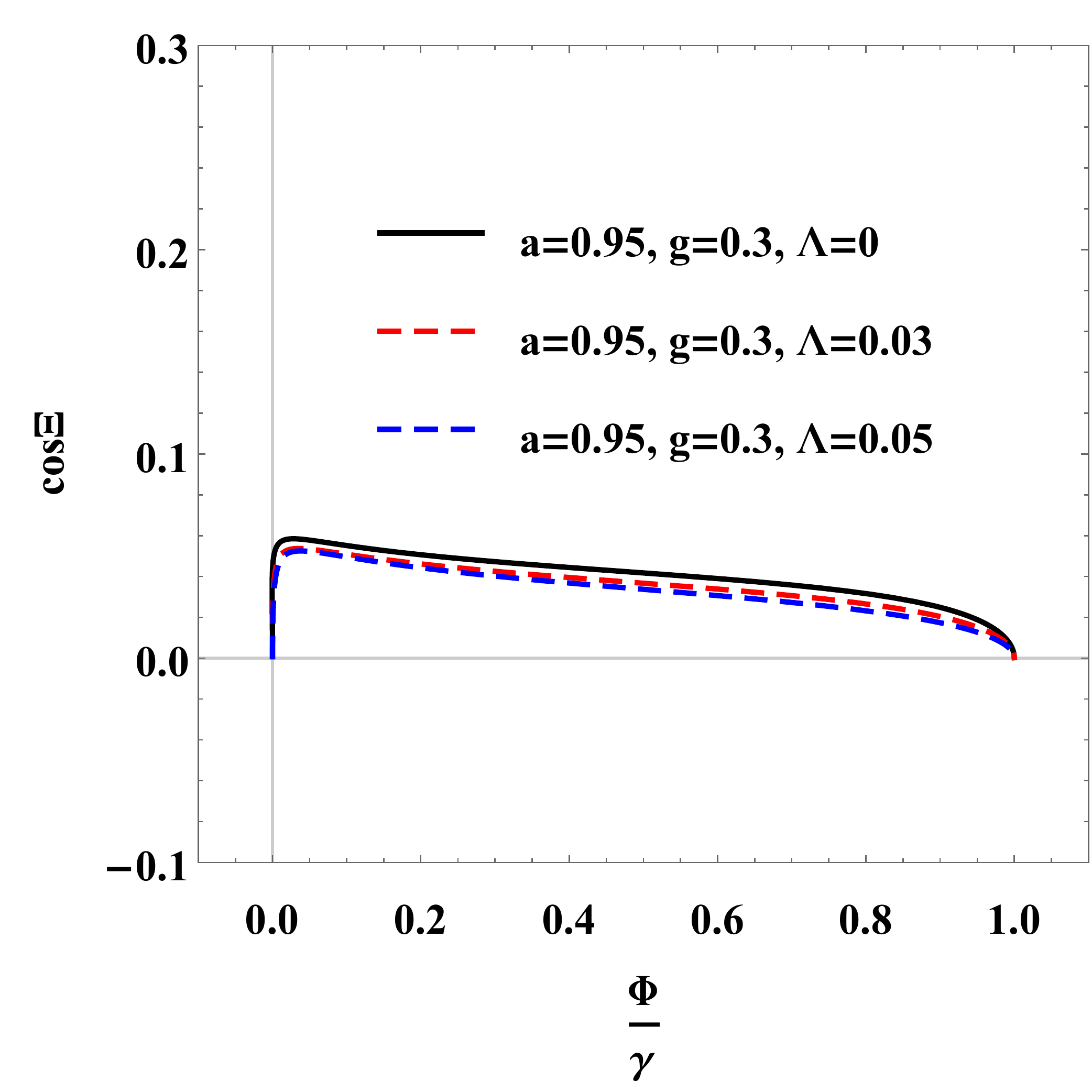}
			\end{minipage}\label{8a}
		}	
		\subfigure[]{
			\begin{minipage}[t]{0.311\textwidth}
				\includegraphics[width=1\textwidth]{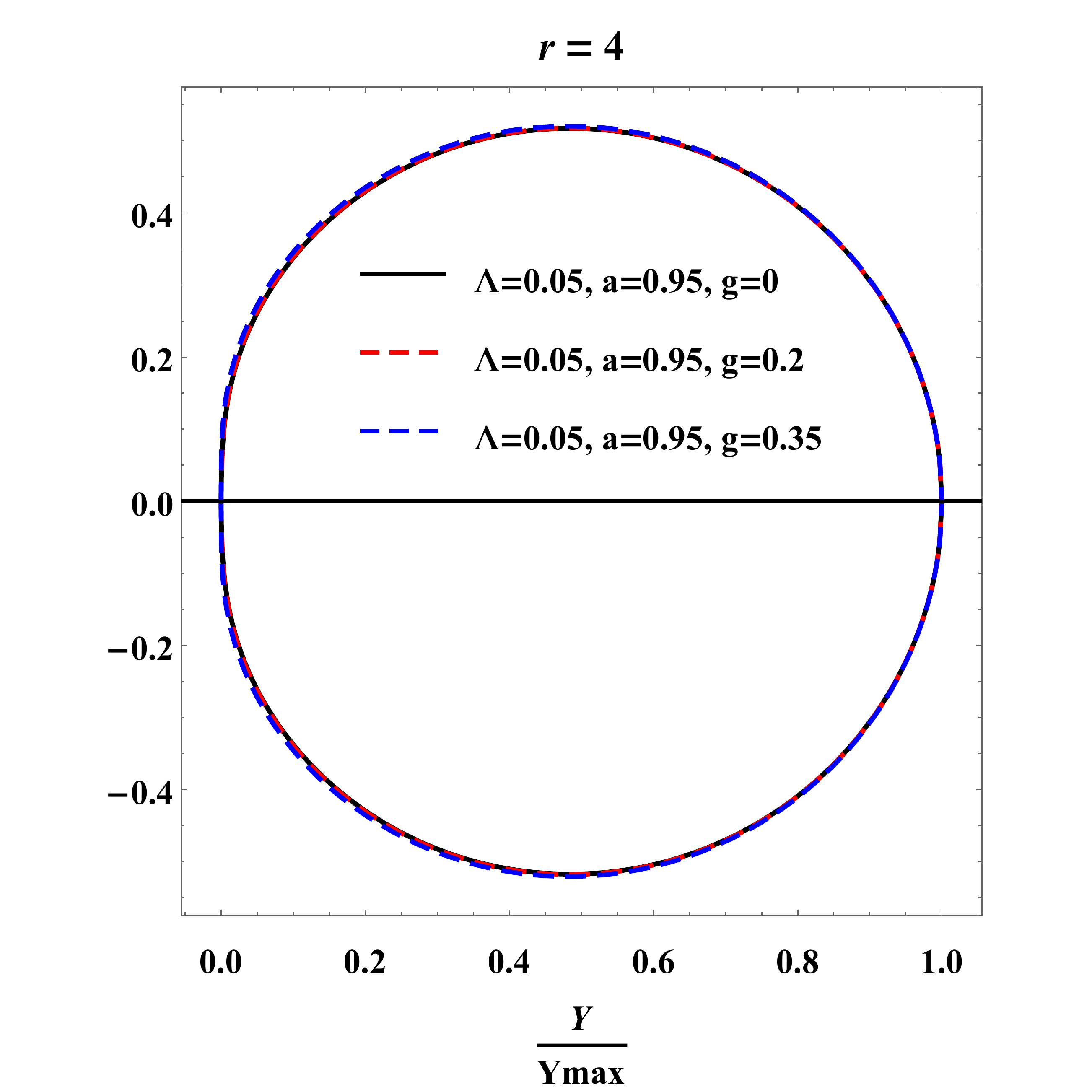}
				\includegraphics[width=1\textwidth]{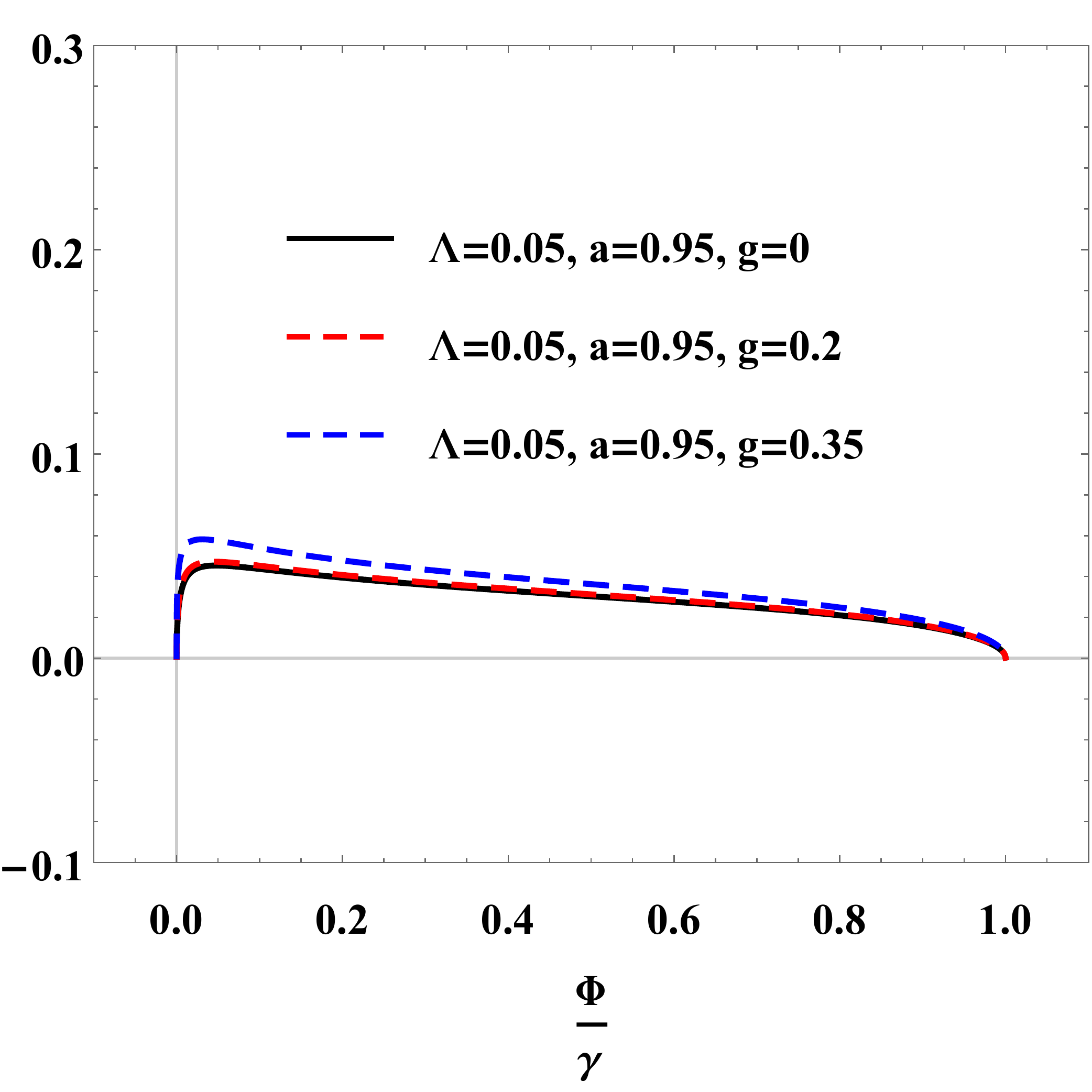}
			\end{minipage}\label{8b}
		}
		\subfigure[]{
			\begin{minipage}[t]{0.311\textwidth}
				\includegraphics[width=1\textwidth]{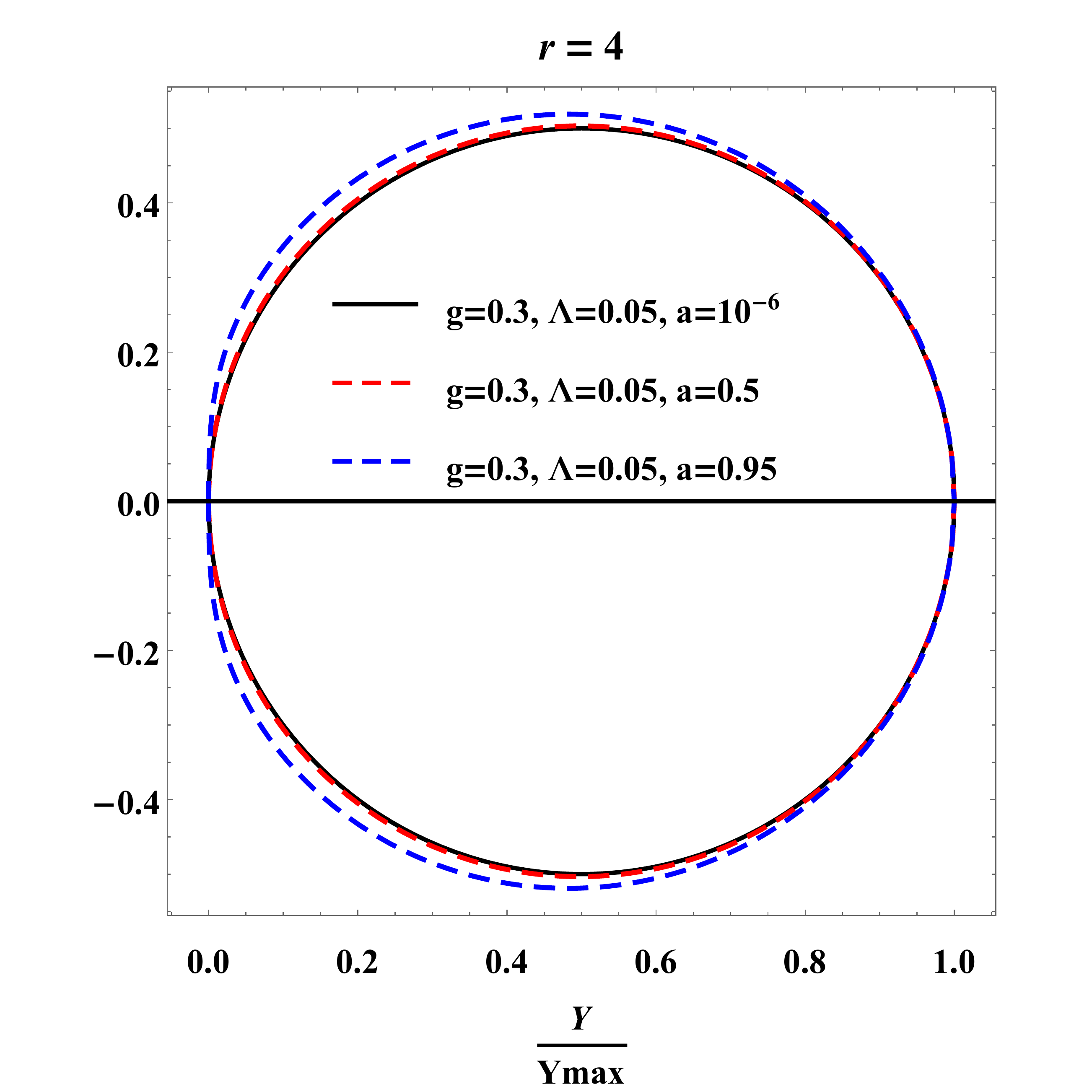}
				\includegraphics[width=1\textwidth]{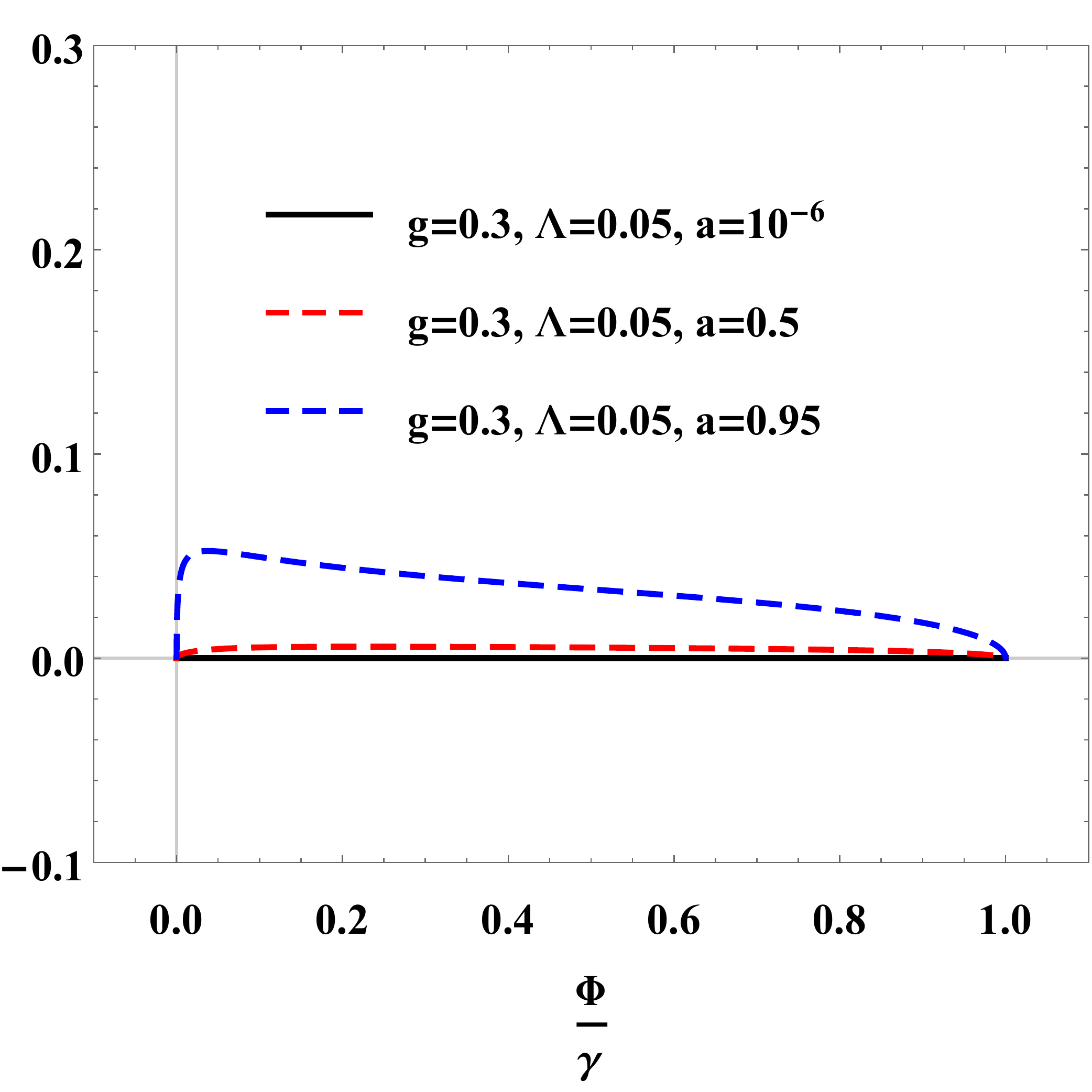}
			\end{minipage}\label{8c}
		}
		\caption{The shape of shadows and corresponding distortion parameters $ \Xi $ as function of $ \frac{\Phi}{\gamma} $ for selected different parameters for observers located at $ r=4 $. Here we set $ M=1 $. (a) Shadows and distortion parameters of rotating Hayward-de Sitter black holes of selected different cosmological constants. (b) Shadows and distortion parameters of rotating Hayward-de Sitter black holes of selected different magnetic monopole charges. (c) Shadows and distortion parameters of rotating Hayward-de Sitter black holes of selected different the spin parameters.}
		\label{fig:8}
	\end{figure}

\begin{figure}[htbp]
	\centering
	\subfigure[]{
		\begin{minipage}[t]{0.22\textwidth}
			\includegraphics[width=\textwidth]{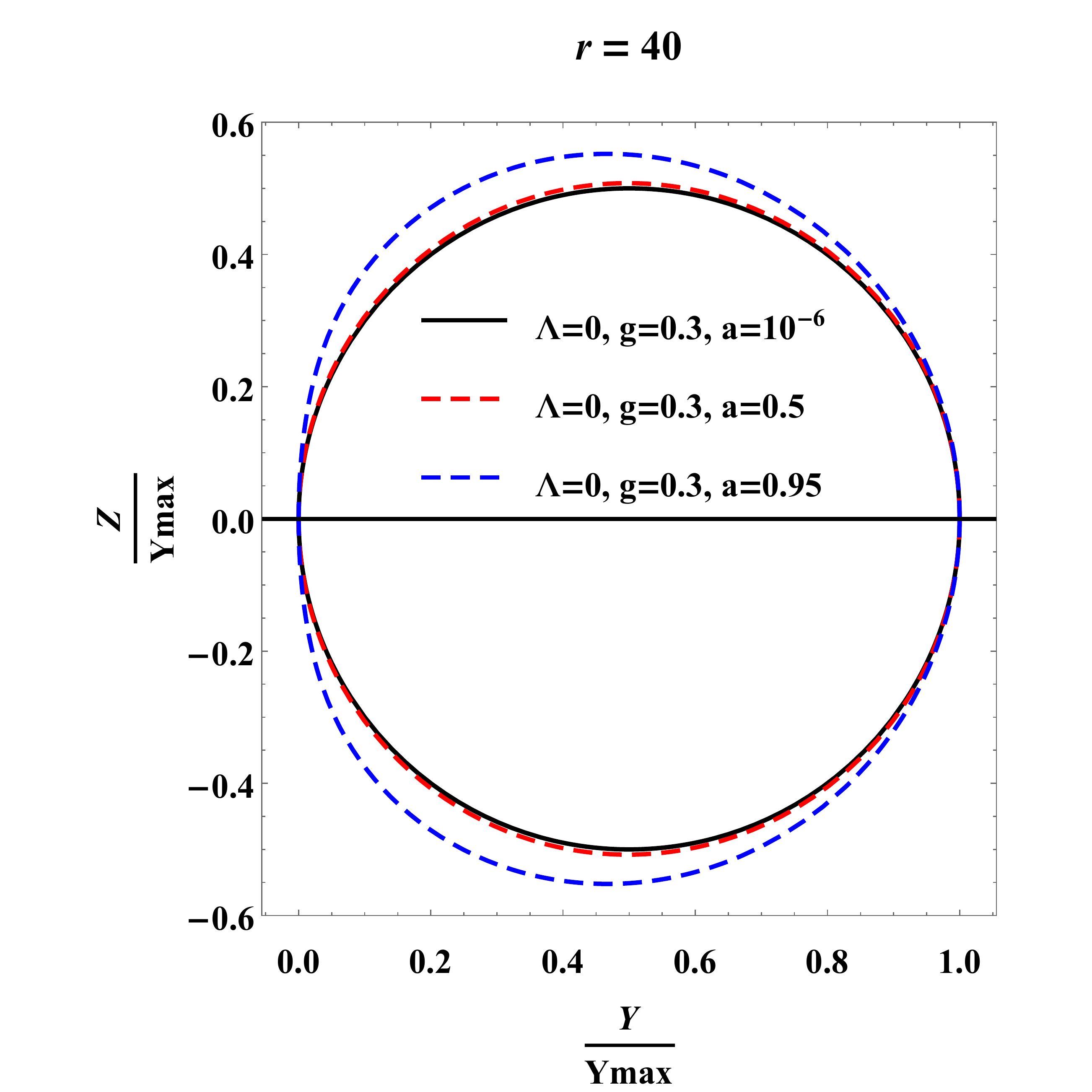}
			\includegraphics[width=\textwidth]{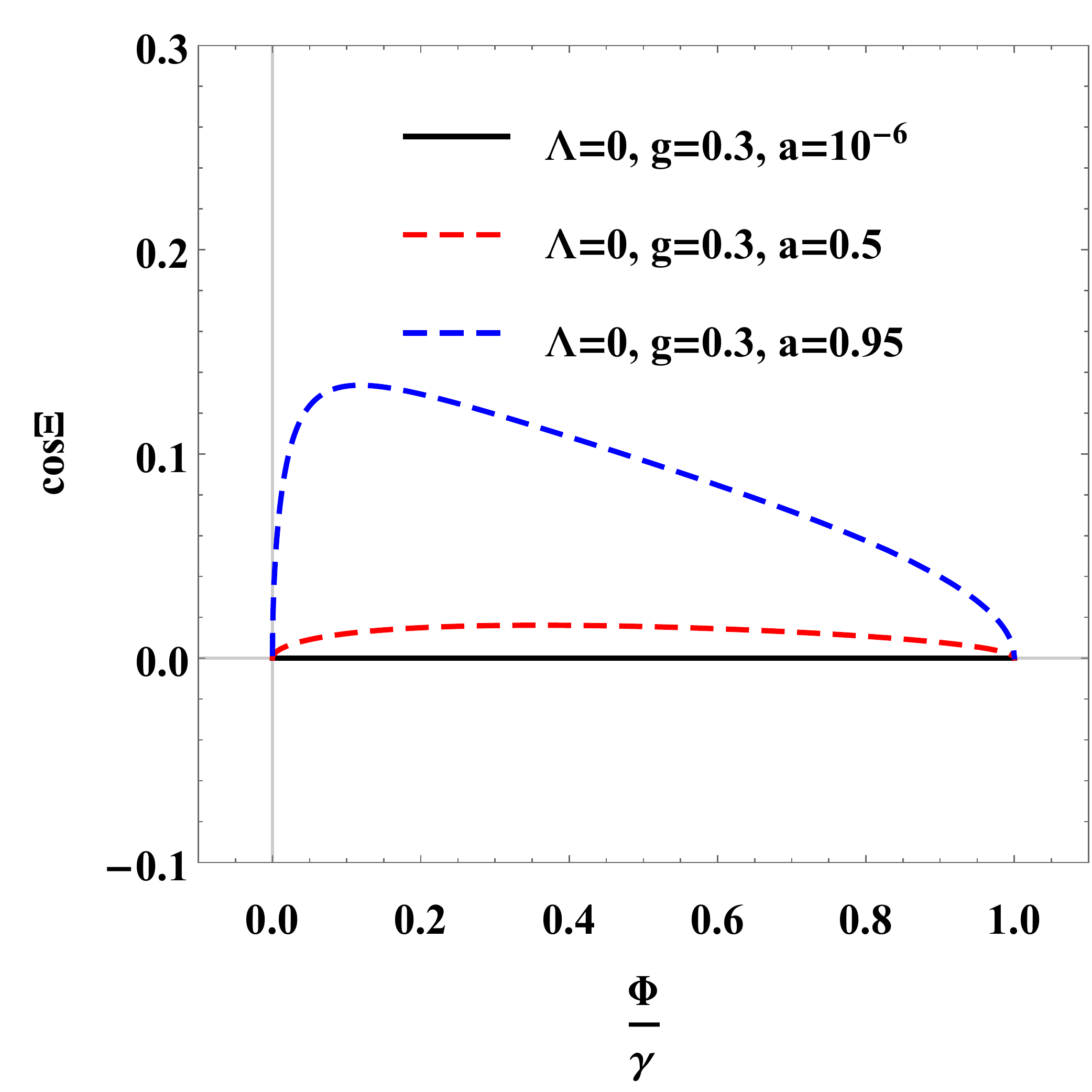}
		\end{minipage}\label{9a}
	}	
	\subfigure[]{
		\begin{minipage}[t]{0.22\textwidth}
			\includegraphics[width=\textwidth]{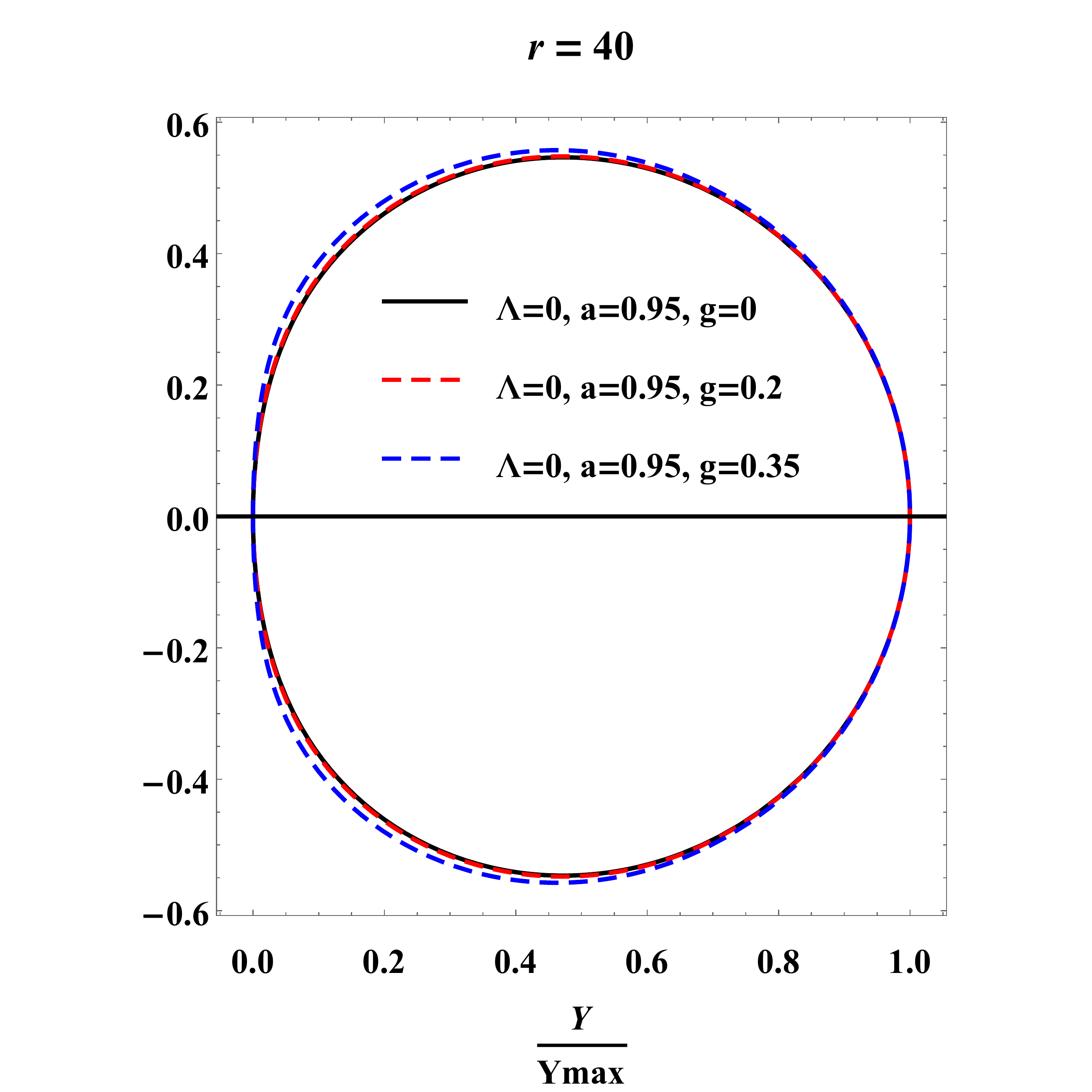}
			\includegraphics[width=\textwidth]{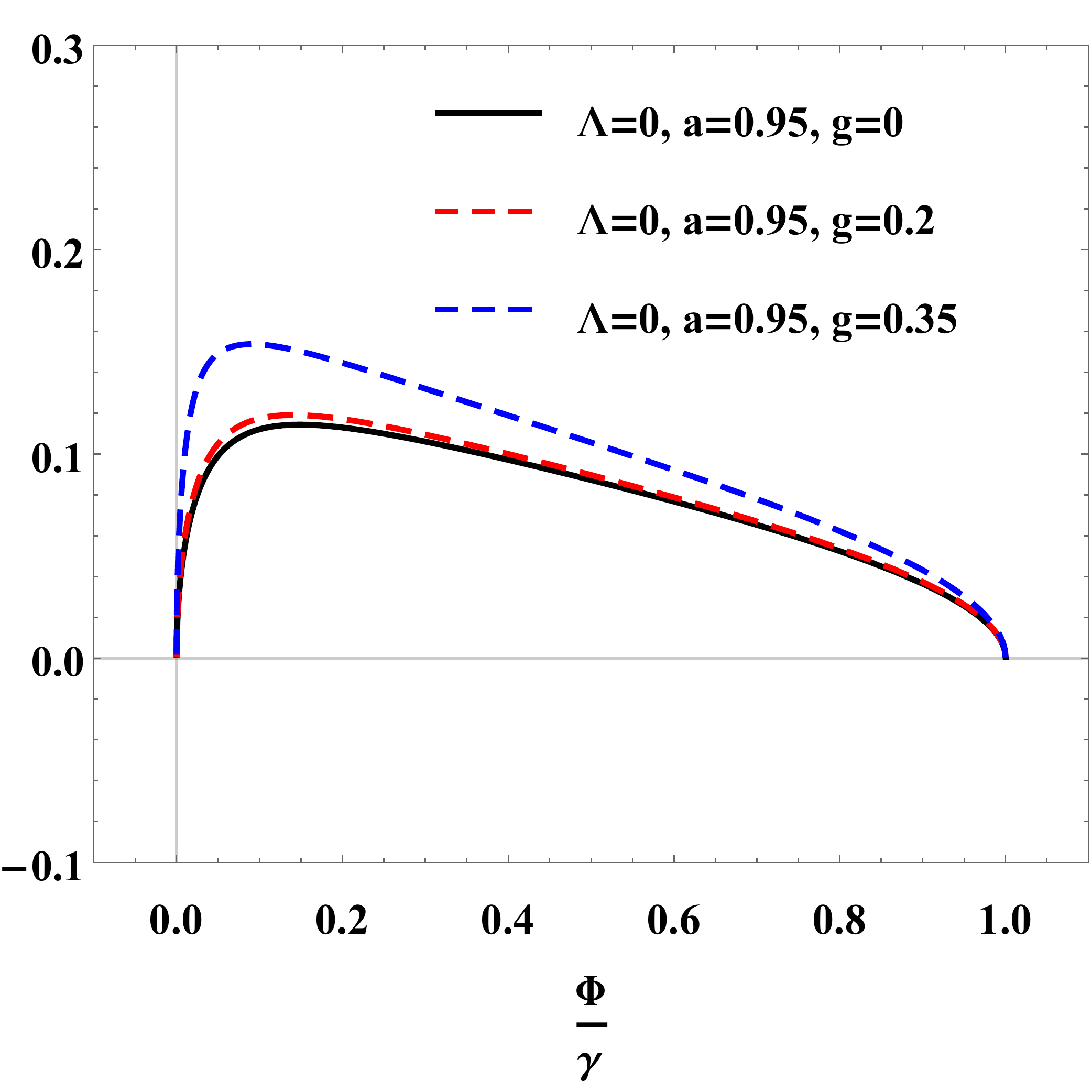}
		\end{minipage}\label{9b}
	}
	\subfigure[]{
		\begin{minipage}[t]{0.22\textwidth}
			\includegraphics[width=\textwidth]{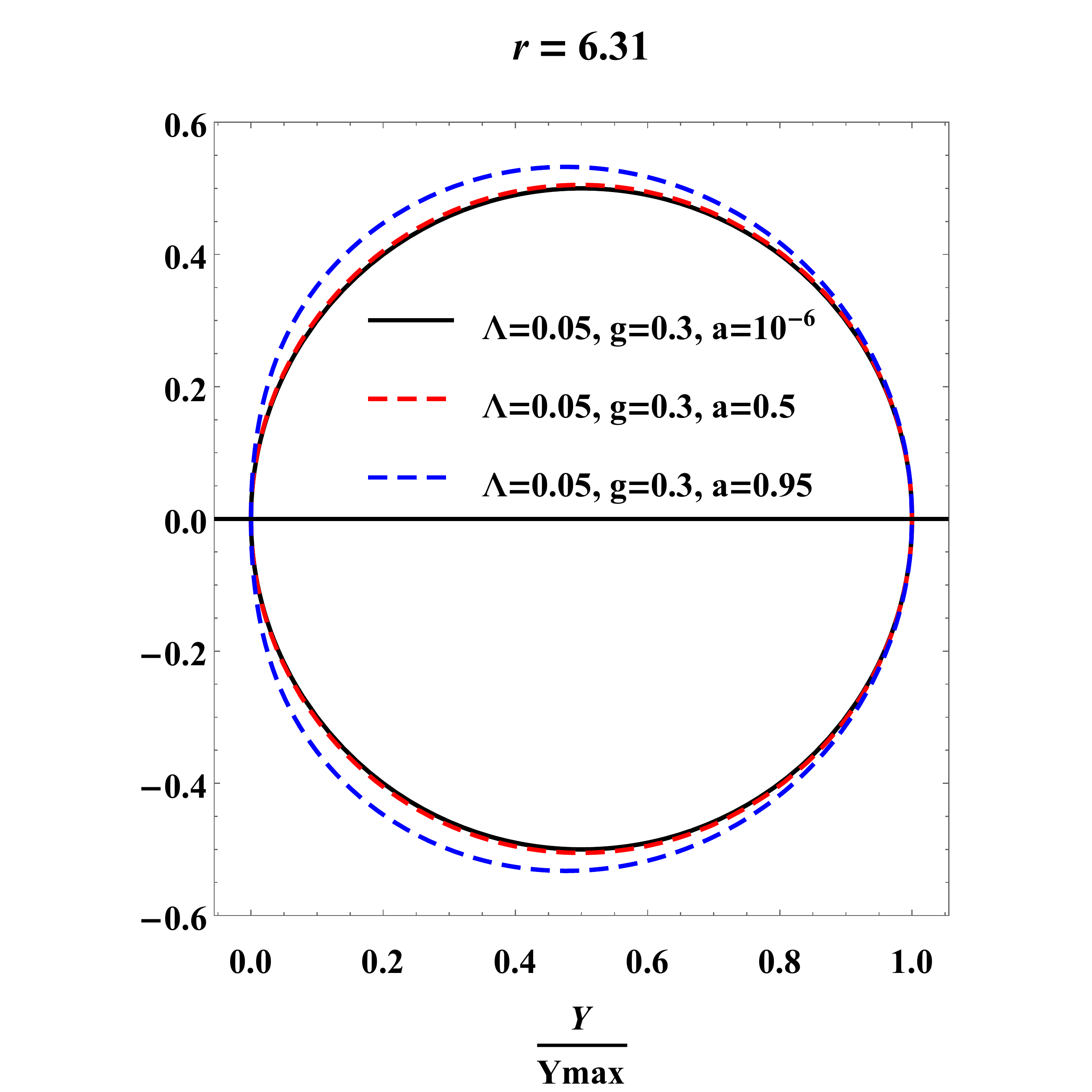}
			\includegraphics[width=\textwidth]{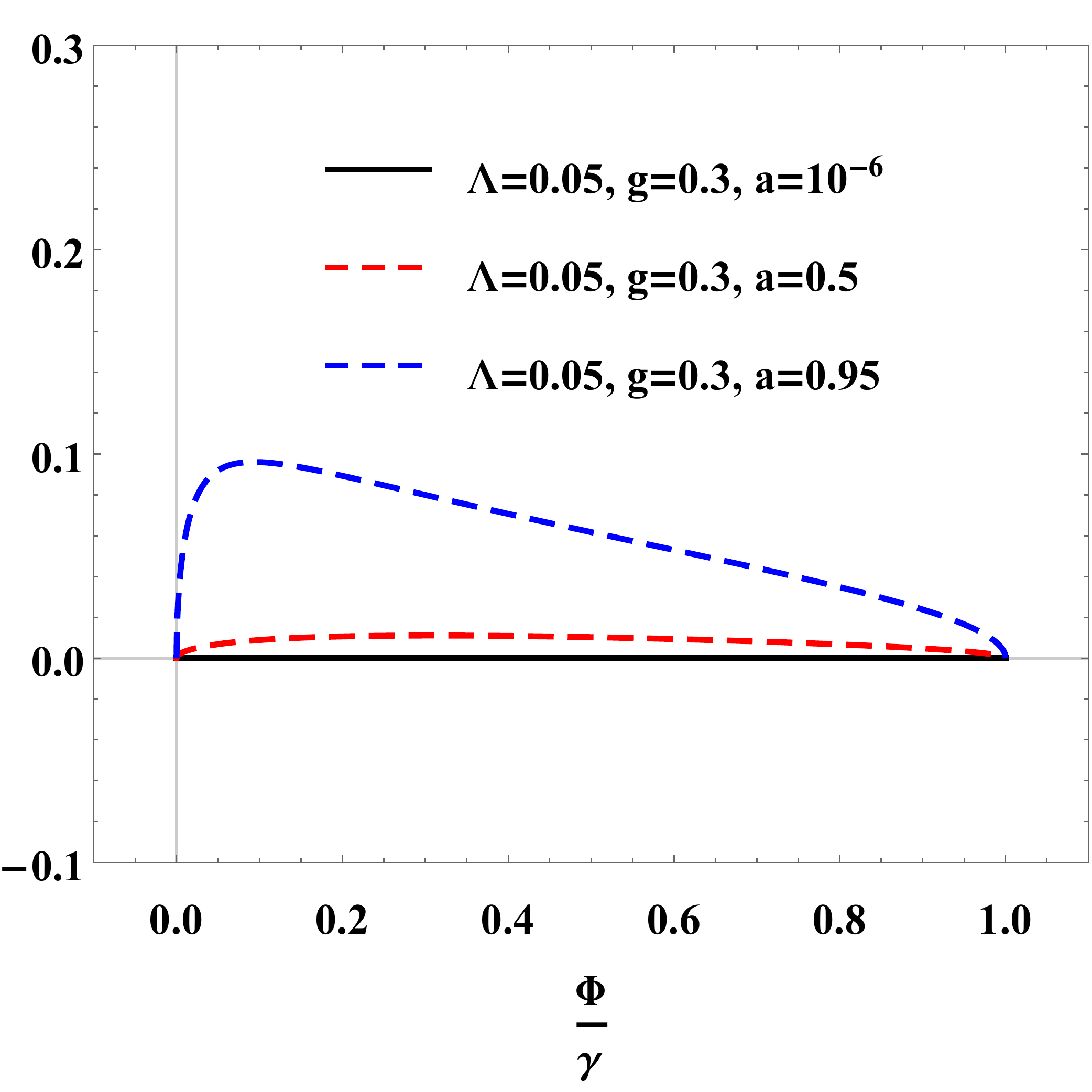}
		\end{minipage}\label{9c}
	}
\subfigure[]{
	\begin{minipage}[t]{0.22\textwidth}
		\includegraphics[width=\textwidth]{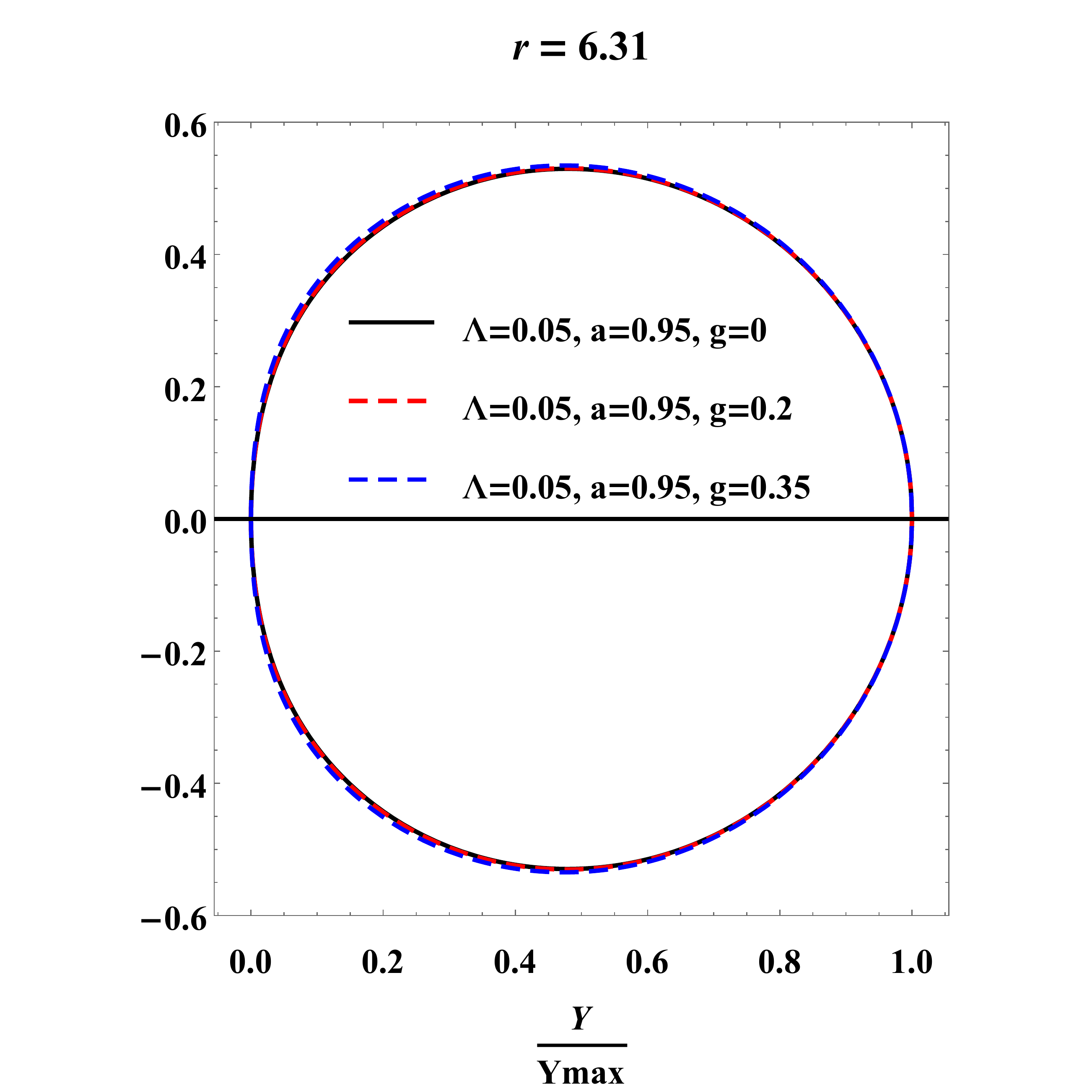}
		\includegraphics[width=\textwidth]{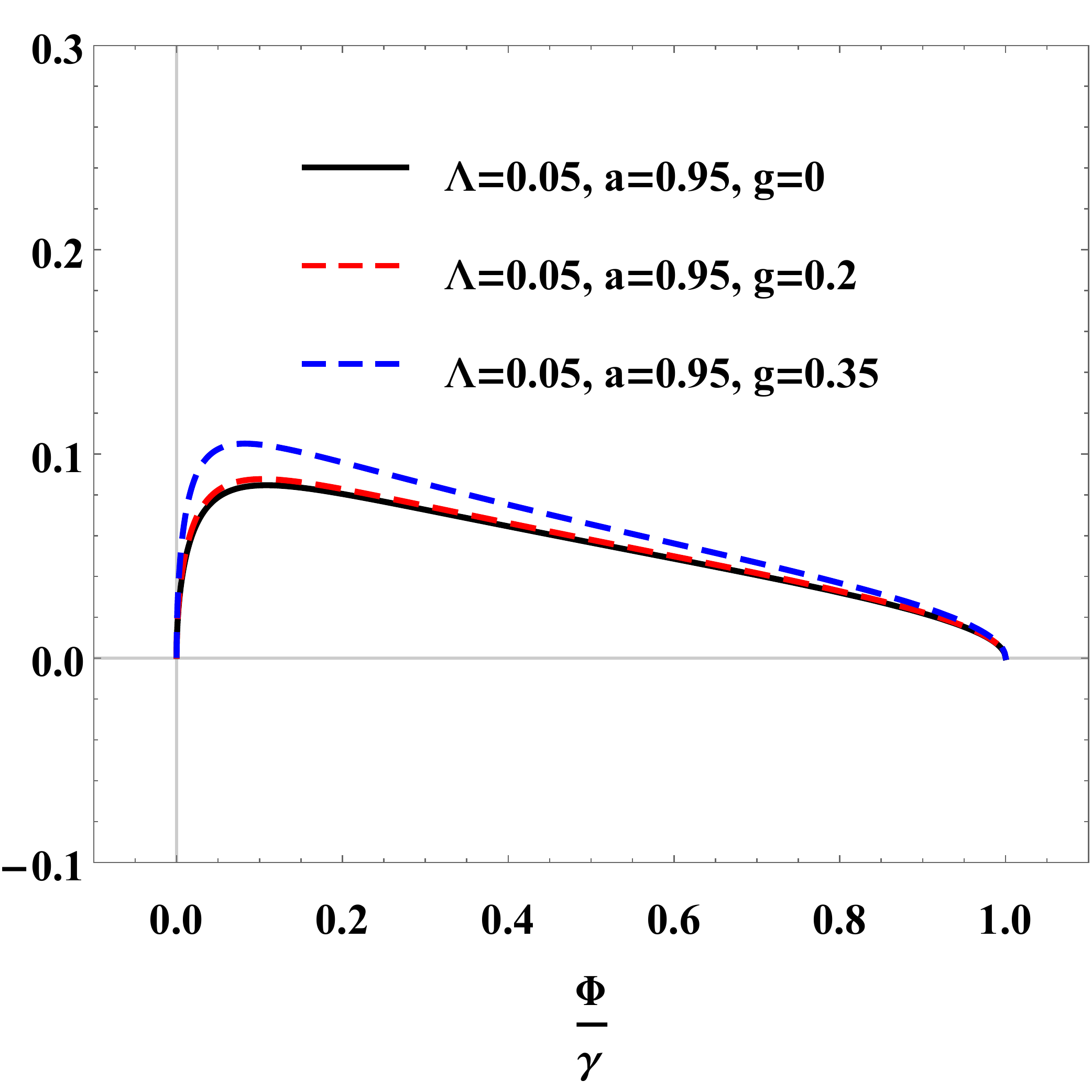}
	\end{minipage}\label{9d}
}
	\caption{The shapes of shadows and corresponding distortion parameters $ \Xi $ as function of $ \frac{\Phi}{\gamma} $ for selected different parameters for distant observers. Here we set $ M=1 $. (a) Shadows and distortion parameters of rotating Hayward black holes of selected different spin parameters for observers located at $ r=40 $. (b) Shadows and distortion parameters of rotating Hayward black holes of selected different magnetic monopole charges for observers located at $ r=40 $. (c) Shadows and distortion parameters of rotating Hayward-de black holes of selected different spin parameters for observers located at $ r=6.31 $. (d) Shadows and distortion parameters of rotating Hayward-de black holes of selected different the magnetic monopole charges for observers located at $ r=6.31 $.  }
	\label{fig:9}
\end{figure}

\begin{figure}[htbp]
	\centering
	\subfigure[]{
		\begin{minipage}[t]{0.475\textwidth}
			\includegraphics[width=\textwidth]{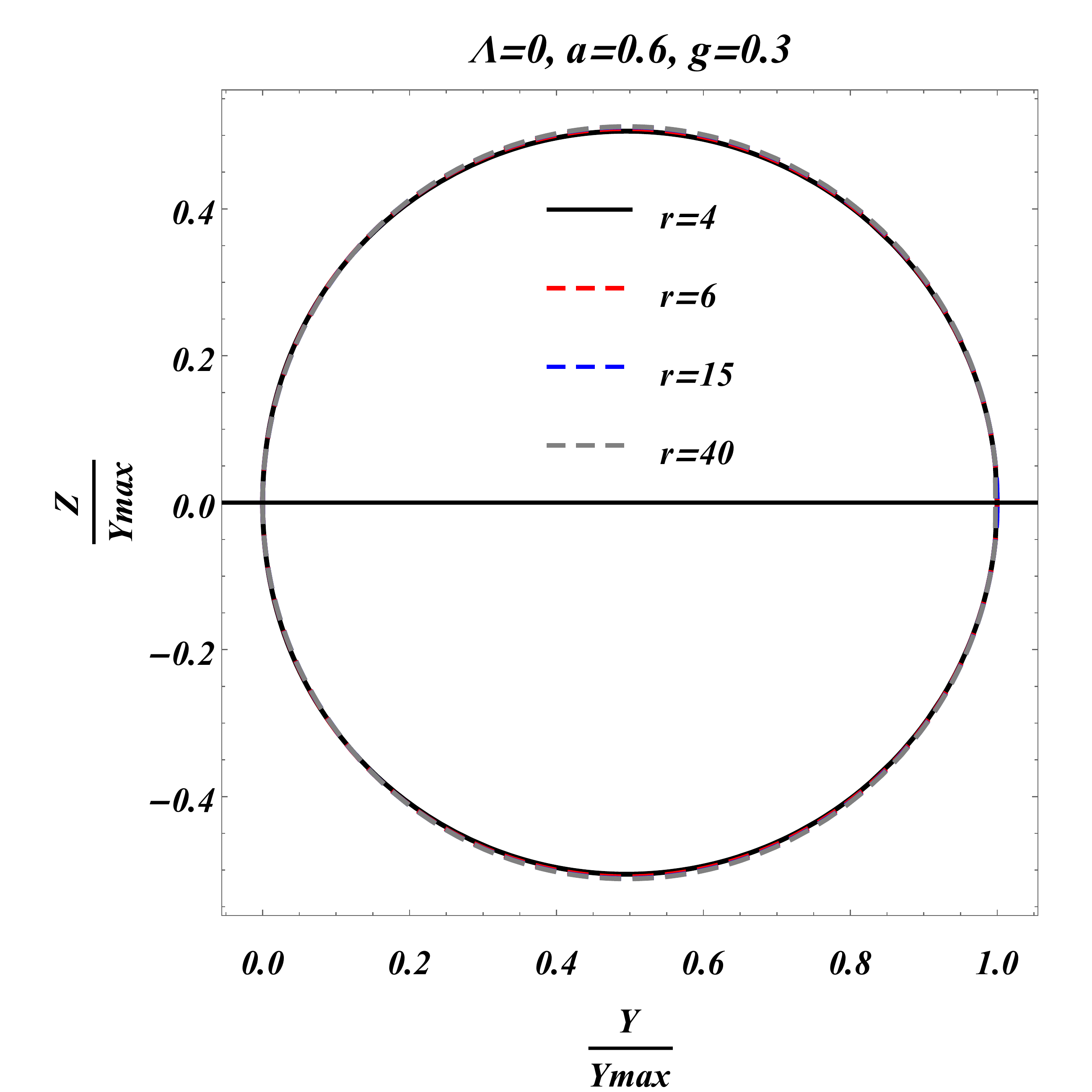}
			\includegraphics[width=\textwidth]{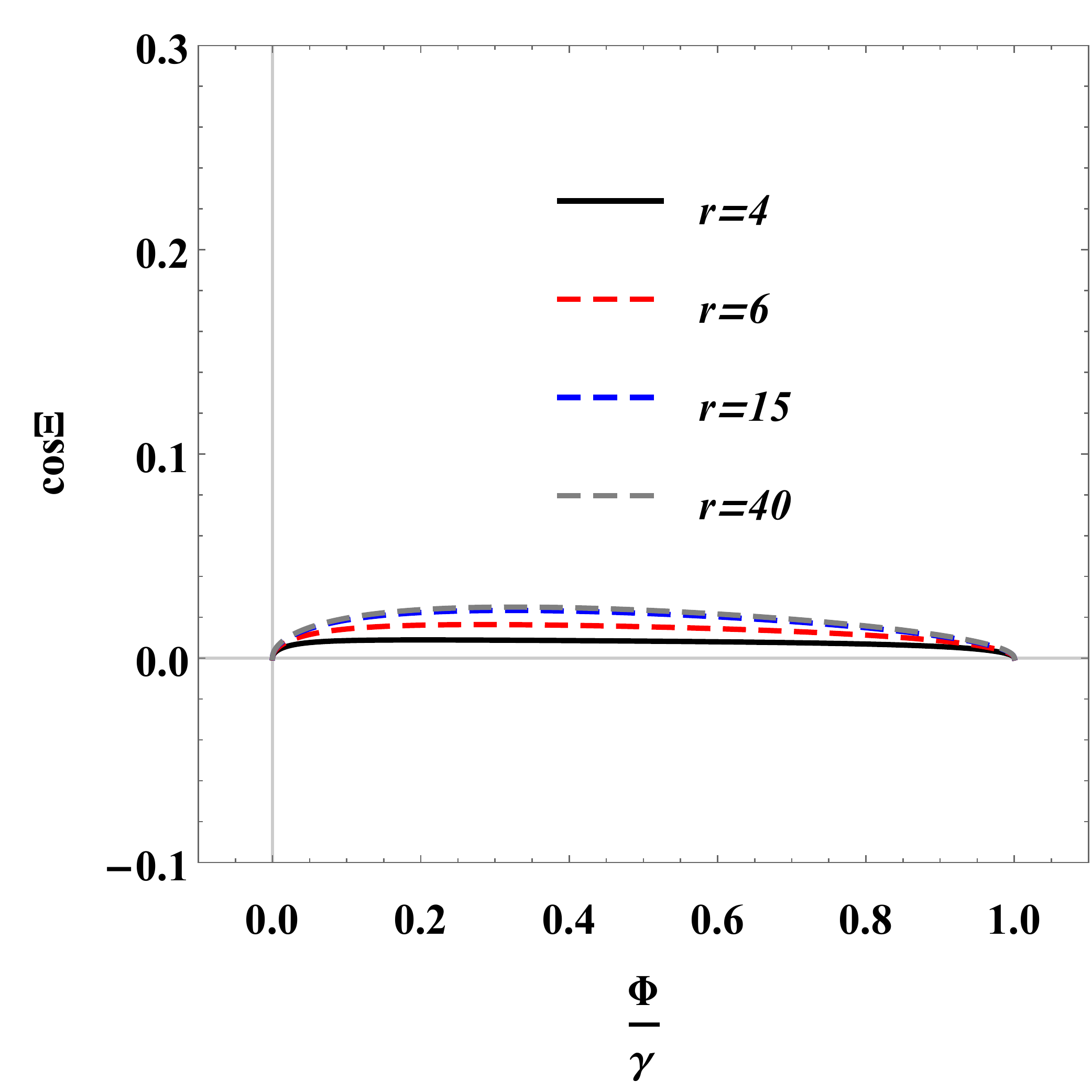}
		\end{minipage}\label{10a}
	}	
	\subfigure[]{
		\begin{minipage}[t]{0.475\textwidth}
			\includegraphics[width=\textwidth]{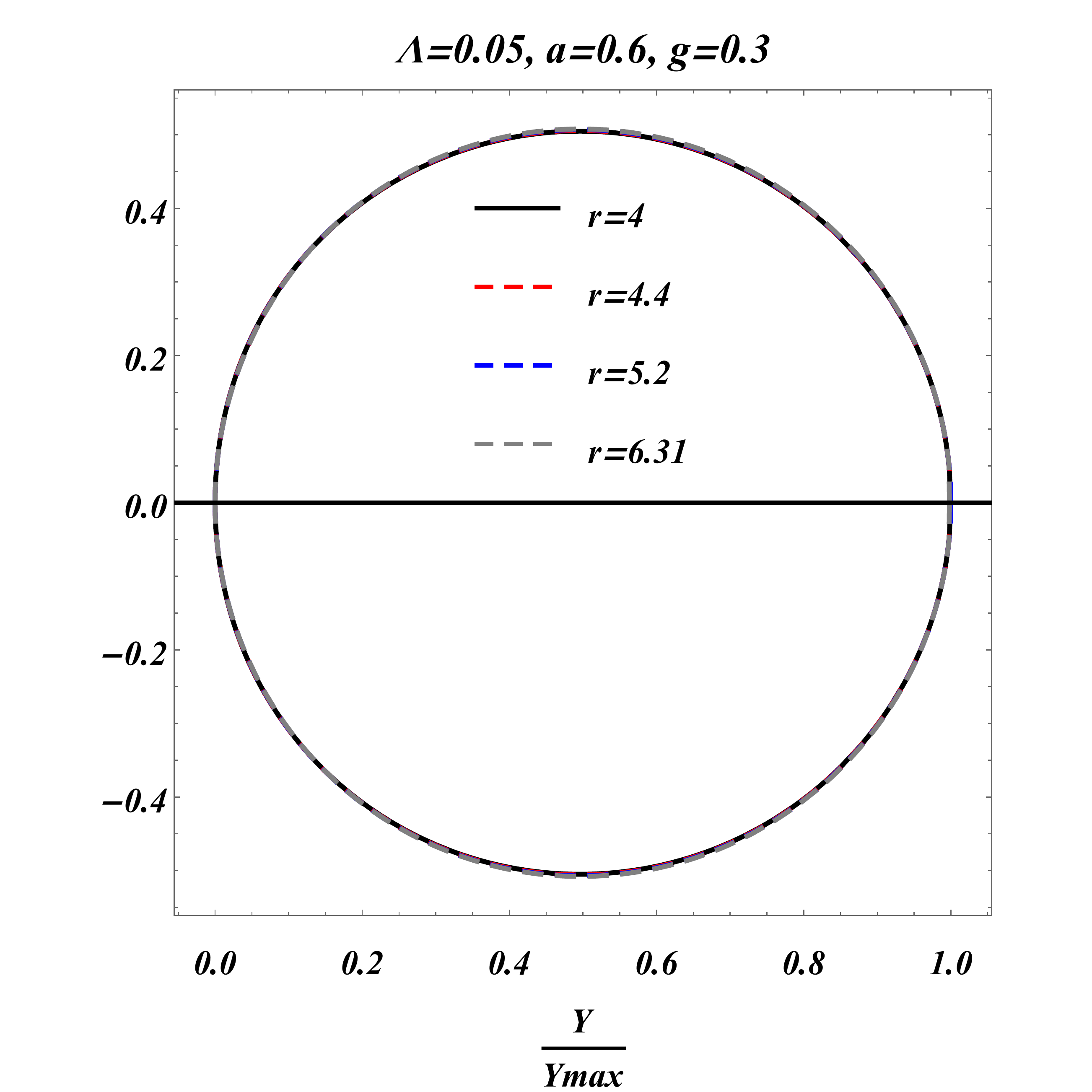}
			\includegraphics[width=\textwidth]{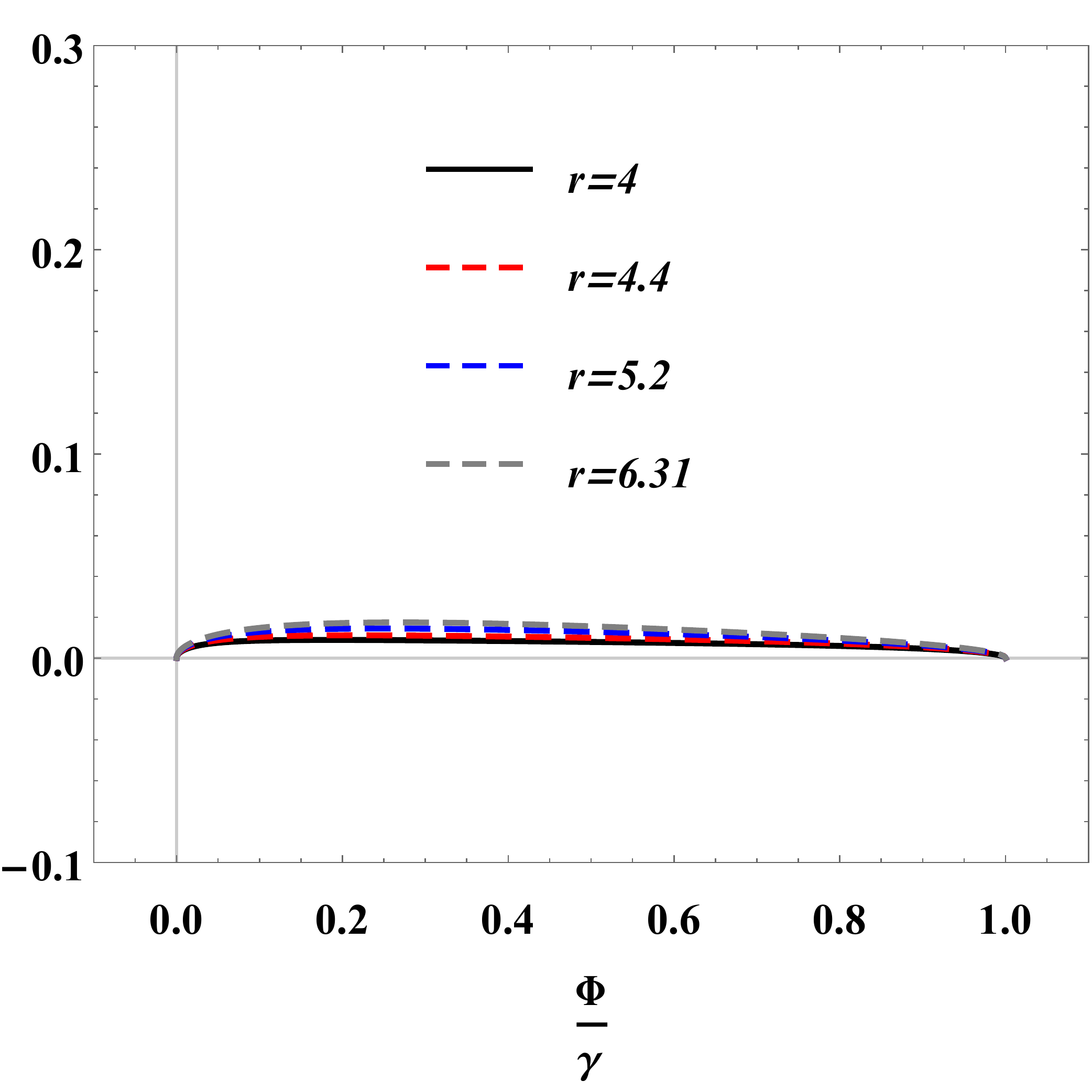}
		\end{minipage}\label{10b}
	}
	\caption{The shapes of shadows and corresponding distortion parameters $ \Xi $ as function of $ \frac{\Phi}{\gamma} $ for observers at selected position $ r $. }
	\label{fig:10}
\end{figure}
 \par
In Fig.~\ref{fig:7}, we set $g=0  $ and get the same results as the Kerr(-de Sitter) black holes in Ref.~\cite{changzhe2}. In Figs.~\ref{fig:8} and~\ref{fig:9}, we scale the shadows appropriately so that the degree of distortion of these shadows can be compared qualitatively. The upper parts of Figs.~\ref{fig:8} and~\ref{fig:9} are the shadows after scaling, with different parameters selected, and the lower parts are the images of the corresponding distortion parameters vary with $ \Phi/\gamma $, which are the quantitative description of the shadow's distortion. In Fig.~\ref{fig:8}, the observers are not far from the photon regions of the black holes, and in Fig.~\ref{fig:9}, the observers are far away from the black holes. It is not difficult to find that the shadow's distortion will decrease as the cosmological constant increases. In contrast, the distortion will increase with an increase of $ g $ or $a $.
\par
In Fig.~\ref{fig:10}, we plot the shapes and distortion parameters of the shadows for observers at different distances from the center of the black hole. It can be seen that the distortion parameter would increase with distance. Through the above discussion, we know that when the parameters $ g $ and $ a $ of rotating Hayward-de Sitter black holes are maximum, and the cosmological constant is zero, the distortion of the shadow is the largest.

\section{CONCLUSIONS AND DISCUSSIONS}\label{sec4}
In this article, we calculated the size and shape of rotating Hayward-de Sitter black hole shadow for static observers at a finite distance in terms of astronomical observables. For $ \theta =0 $, the shadow's boundary is a perfect circle, but for $ \theta =\frac{\pi}{2} $, the shadow's boundary will be distorted. To quantitatively describe the distortion of the shadows, we plotted the distortion parameter affected by the black hole's parameters, quantifying the distortion of the shape from circularity.   We found that no matter which parameter increases, the size of the shadow will shrink. At the same distance, a Schwarzschild black hole has the largest shadow. Furthermore, the parameters $ g $ and $ a $ of rotating Hayward-de Sitter black holes are maximum, and the cosmological constant is zero, the distortion of the black hole shadow is the largest, and the distortion parameter would increase with distance.
\par
We only considered static observers fixed at inclination angle $ \theta =0 $ and $ \theta =\frac{\pi}{2} $, but this method is suitable for arbitrary observers. Studying the shadows of black holes is an important way for studying the properties of black holes, from which one can obtain rich information about space-time geometry.

%\newpage  
\section*{Conflicts of Interest}
  The authors declare that there are no conflicts of interest regarding the publication of this paper.

\section*{Acknowledgments}
  We would like to thank the National Natural Science Foundation of China (Grant No.11571342) for supporting us on this work.

\end{document}